\documentclass[11pt]{article}
\usepackage{amsmath,amsthm,amssymb}
\usepackage{graphicx}
\usepackage{enumerate}
\theoremstyle{plain}
\newtheorem{thm}{Theorem}
\newtheorem{lem}[thm]{Lemma}
\newtheorem{prop}[thm]{Proposition}
\newtheorem{rem}[thm]{Remark}

\theoremstyle{definition}
\newtheorem{rem*}{Remark}
\newtheorem*{rems*}{Remarks}
\newtheorem*{def*}{Definition}

\providecommand{\R}{\mathrm}

\newcommand{\dd}{\mathrm{d}}
\newcommand{\ee}{\mathrm{e}}
\newcommand{\ii}{\mathrm{i}}
\renewcommand{\Im}{\operatorname{Im}}

\renewcommand{\Re}{\operatorname{Re}}

\DeclareMathOperator{\diag}{diag}

\begin{document}
\title{Robin boundary condition and shock problem for 
the focusing nonlinear Schr\"odinger equation}
\author{Spyridon Kamvissis$^1$, Dmitry Shepelsky$^2$ and Lech Zielinski$^3$}
\date{}
\maketitle

\begin{abstract}
We consider the initial boundary value (IBV) problem
for the focusing nonlinear Schr\"odinger equation in the
quarter plane $x>0,t>0$ in the case of periodic initial data (at $t=0$)
and a Robin boundary condition at $x=0$. Our approach is based on the simultaneous
spectral analysis of the  Lax pair equations combined with symmetry
considerations for the corresponding Riemann-Hilbert problems.
A connection between the original IBV problem and an associated initial value (IV)
problem is established.
\end{abstract}

\section{Introduction}
\setcounter{equation}{0}

Any adaptation of the inverse scattering transform (IST) method
to the study of initial boundary value (IBV)
problems for nonlinear evolution equations possessing a Lax pair representation
(called \textit{integrable}) faces a major problem: the evolution of the spectral data
require knowledge of an ’’excessive'' amount of  boundary values: 
they cannot all be given as boundary conditions for a 
well-posed IBV problem. However, certain particular classes of boundary conditions, 
under which the 
IBV problem remains completely integrable, i.e. solving it reduces to solving a series of well-posed 
linear problems, can be specified. 

In the case of  decaying initial data, these classes (for various nonlinear equations) are known to 
be related to an appropriate continuation,
based on the B\"acklund transformation,
 of the given initial data to the whole axis,
which  reduces the study of the IBV problem to the study of the associated initial value (IV) problem.
Although this was already realized at the beginning of 1990s  \cite{BT91, F89, Sk, T88,   T91},
the continuation approach has been adjusted to the modern, Riemann--Hilbert framework,
only recently \cite{DP10, IS12}. A primary importance of this framework is that it makes possible
the rigorous   study of the large-time asymptotics via  the nonlinear steepest descent method
\cite{DZ93}.

An alternative approach to the IBV problems with these special boundary conditions
(called \textit{linearizable}) stems from a general approach to IBV problems for integrable nonlinear
equations  originated by Fokas  \cite{F93}. The basic idea of this approach consists in  considering the both
linear equations of the associated Lax pair as spectral problems, each generating respective sets of spectral 
functions. Consequently, when applied to problems on \text{infinite} time intervals (say, $t\in (0,\infty)$),
this approach requires  defining spectral functions associated with the boundary values on these intervals;
and thus an information on the large-$t$ behavior of these boundary values is required.
The problem here is that, as we have already mentioned above, only a part of  boundary values (determining the spectral functions)
can be given as boundary conditions;  the behavior of the remaining part is  unknown
and thus has to be provided as an a priori information.

In \cite{IS12} it has been observed that the Riemann--Hilbert problem obtained
in \cite{FIS05} and \cite{FK04} for linearizable boundary conditions  under  a priori assumption
that the boundary data rapidly decay as $t \to \infty$ provides the solution of the associated IBV problem
without any such assumption, because one can prove independently the validity of the required initial and
boundary conditions. The proof goes actually back to the late 1980s works
on algebro-geometric  solutions of integrable PDEs and is based on 
the symmetry considerations for the \textit{deformed} Riemann--Hilbert problem.

All the studies mentioned above are related to the problems with decaying (as $x\to\infty$) initial data.
On the other hand, problems (particularly, initial value problems on the whole line) with
non-decaying initial data attract a growing interest. 
As in the case of the decaying boundary conditions, the Riemann--Hilbert approach allows a detailed study of
the large-time behavior of solutions. In \cite{BKS11}, the initial value (IV) problem for the focusing NLS with generalized ''step-like''
initial data (vanishing for $x<0$ and periodic in $x$ 
($u(x,0)= \alpha\exp (-2 \ii \beta x)$) for $x>0$) is considered and the 
large-time behavior of the solution is presented. 

In \cite{BV07}, the initial value problem for the 
 NLS equations is considered with oscillatory initial data $\alpha \exp(-2 \ii \beta |x|)$ with  $\beta>0$,
which can be viewed as the problem of collision of two plane waves, $ \alpha \exp(-2 \ii \beta x + \ii \omega t))$ and 
 $\alpha  \exp(2 \ii \beta x + \ii \omega t)$ resembling the Toda shock problem \cite{VDO}. In \cite{BV07}, three regions in space-time
 with qualitatively different asymptotic behavior of the solution are described, showing, in particular, that for any $x$ fixed,
 the large-$t$ behavior is described in terms of single phase theta functions.
 
 Notice that in \cite{BV07}, the considered initial data are even w.r.t. $x$, and so is the solution of the IV problem for any $t$.
 Particularly, this solution satisfies the condition $u_x(0,t)=0$ for all $t>0$.
 Saying differently, the solution $u(x,t)$ of the IV problem in  \cite{BV07} is the even continuation
 of the solution of the initial boundary value problem:
 \begin{equation} \label{ibv-n}
\begin{aligned}
   & \ii u_t + u_{xx} + 2|u|^2 u =0,  & x>0, t>0, \\
& u(x,0)=\alpha \ee^{-2\ii \beta x}, & x> 0, \\
& u_x(0,t)=0, \quad & t> 0, 
 \end{aligned}
\end{equation}
where $\alpha>0$ and $\beta>0$ are  constants.
 
 In the present paper, we consider the initial boundary value problem 
 for the focusing nonlinear Schr\"odinger equation with oscillatory initial data
 and the Robin boundary condition:
\begin{equation} \label{ibv}
\begin{aligned}
   & \ii u_t + u_{xx} + 2|u|^2 u =0,  & x>0, t>0, \\
& u(x,0)=u_0(x), & x> 0, \\
& u_x(0,t)+q u(0,t)=0, \quad & t> 0, 
 \end{aligned}
\end{equation}
where 
$$
u_0(x)-\alpha \ee^{-2\ii \beta x}\to 0  \quad \text{as}\ \ x\to+\infty.
$$
Here
$\alpha>0$ whereas $\beta\ne 0 $ and $q\ne 0$ are real constants which can be positive or negative.
To be more specific, a special attention will be paid to the case of pure oscillatory initial data:

$$
u_0(x)=\alpha \ee^{-2\ii \beta x} \quad \text{for all}\  x>0.
$$

Our approach follows the lines of the general approach \cite{BFS04,FIS05} to IBV problems 
based on the use of the associated Riemann--Hilbert problem 
(involving the spectral functions 
associated with the initial condition and the Robin parameter $q$)
and consists of the following main steps:
(i) 
provide a family of the Riemann--Hilbert 
problems (parametrized by $x$ and $t$) such that the solution $u(x,t)$ of (\ref{ibv})
is expressed in terms of the solutions of these problems;
(ii)
 prove that  $u$ satisfies the initial condition $u(x,0)=u_0(x)$;
(iii) prove that $u$ satisfies the boundary condition $u_x(0,t)+q u(0,t)=0$.
In our approach, item (i) follows the 
general framework of the simultaneous spectral 
analysis of the Lax pair equations 
whereas the proof of (ii) and (iii) are based on appropriate deformations of the original 
RH problem. 

When proving (iii) we deform the RH problem in such a way that the 
resulting RH problem can be viewed as a RH problem associated with an IV problem (on the whole line $-\infty<x<\infty$)
whose restriction to the half-line $x>0$ gives the solutions of the IBV problem (\ref{ibv}).
Particularly, 
in the case $u_0(x)=\alpha \ee^{-2\ii \beta x}$ for $x>0$, we will show that 
$u(x,0) = \alpha \ee^{\ii\phi} \ee^{2\ii \beta x}$ for 
 $x<0$,  for a certain real phase $\phi$,
and thus the IV problem itself can be viewed as a generalized shock problem.

On the other hand, the RH problem formulation allows the detailed study of the large-time asymptotics of the solution
of the constructed IV problem as well as of the solution of the original IBV problem 
(\ref{ibv}). 
Particularly, the impact of the Robin parameter to the large-$t$ behavior of the solution of the IBV (or IV) problem at $x=0$,
observed in the case of the decaying initial conditions, see \cite{IS12}, is present in the case of oscillatory initial conditions as well,
namely: in the case $q>0$, the discrete spectrum associated with the IV problem is always non-empty
 on the imaginary axis, and this
generates a breather-type (stationary soliton) oscillatory behavior. Moreover, this breather dominates the large-$t$
behavior of the solution along the $t$-axis in the case  $\beta<0$ (''rarefaction wave'').

\section{The Riemann--Hilbert formalism for the IBV problem with oscillatory initial data}
\setcounter{equation}{0}

In order to adapt 
the Riemann--Hilbert formalism to 
IBV problems on the half-line $x\ge 0$ for the NLS equation \cite{FIS05}
to the case of non-decaying initial data, we need the eigenfunctions and spectral functions
associated with such data \cite{BV07, BKS11}.

Recall that the focusing NLS equation 
\begin{equation}\label{nls}
\ii u_t + u_{xx} + 2|u|^2 u =0
\end{equation}
 is the compatibility condition
of two linear equations (Lax pair) \cite{ZS71}:
\begin{equation}\label{Lax1}
 \Psi_x+\ii k  \sigma_3 \Psi =U \Psi
\end{equation}
with 
\begin{equation}\label{U}
 U=\begin{pmatrix}
       0 & u \\ -\bar u & 0
      \end{pmatrix}
\end{equation}
and 
\begin{equation}\label{Lax2}
 \Psi_t+2\ii k^2  \sigma_3 \Psi =V \Psi
\end{equation}
with 
\begin{equation}\label{V}
 V=\begin{pmatrix}
       \ii |u|^2 & 2ku + \ii u_x \\ -2k\bar u+\ii \bar u_x & -\ii |u|^2
      \end{pmatrix}.
\end{equation}

Let $u^p(x,t):=\alpha \ee^{-2\ii \beta x +2\ii\omega t}$, where $\omega = \alpha^2-2\beta^2$,
 be the plane-wave solution of the 
NLS equation satisfying the initial condition $u^p(x,0) = \alpha \ee^{-2\ii \beta x}$.
Let $U^p(x,t):=\begin{pmatrix}
       0 & u^p(x,t) \\ -\overline{ u^p(x,t)} & 0
      \end{pmatrix}$ 
      and let $\Psi^p(x,t,k)$ be a solution of the Lax pair equations (\ref{Lax1}) and (\ref{Lax2})
      with $u\equiv u^p$:
\begin{equation}\label{phi-p}
\Psi^p(x,t,k) = \ee^{\ii(\omega t -\beta x)\sigma_3}{\cal E}(k)\ee^{(-\ii X(k) x -\ii \Omega(k) t)\sigma_3},
\end{equation}      
where 
\begin{equation}\label{E0}
{\cal E}(k) = \frac{1}{2}\begin{pmatrix}
\nu(k)+\frac{1}{\nu(k)} & \nu(k)-\frac{1}{\nu(k)} \\
\nu(k)-\frac{1}{\nu(k)} & \nu(k)+\frac{1}{\nu(k)}
\end{pmatrix}
\end{equation}
with
\begin{equation}\label{x-om-nu}
\begin{aligned}
\nu(k) & = \left(\frac{k-\beta-\ii\alpha}{k-\beta+\ii\alpha}\right)^{1/4}, \\
X(k) & = \sqrt{(k-\beta)^2+\alpha^2}, \\
\Omega(k) & = 2(k+\beta)X(k).
\end{aligned}
\end{equation}
The functions $\nu(k)$, $X(k)$ and $\Omega(k)$ are defined for $k\in {\mathbb C}\setminus \gamma$,
with 
the branch cut 
$$
\gamma=\{k\equiv k_1+\ii k_2: k_1=\beta, k_2\in (-\alpha, \alpha)\},
$$
in such a way that $\nu(k)=1+o(1)$, $X(k)=k-\beta +o(1)$, and $\Omega(k)=2k^2+\omega+o(1)$ as $k\to\infty$.

Assuming that $u(x,t)$ satisfies (\ref{nls}) for $x>0$ and $0\le t<T$ with some $T<\infty$
and that $u(x,t)-u^p(x,t)\to 0$ as $x\to +\infty$ (for all $0\le t<T$) fast enough,
define the solutions $\Psi_j(x,t,k)$, $j=1,2,3$ of (\ref{Lax1})--(\ref{V})
as follows: 
$$
\Psi_j(x,t,k):=\Phi_j(x,t,k)\ee^{(-\ii k x -2\ii k^2 t)\sigma_3}\quad \text{ for}\ j=1,2
$$
$$
\Psi_3(x,t,k):=\Phi_3(x,t,k)\ee^{(-\ii X(k) x -\ii \Omega(k) t)\sigma_3},
$$
where $\Phi_j$ solve the integral equations 
\begin{subequations}   \label{phi}
\begin{align}
\Phi_1(x,t,k)&=E -\ee^{-\ii kx\sigma_3}\int_t^T \ee^{-2\ii k^2 (t-\tau)\sigma_3}
    V(0,\tau,k) \Phi_1(0,\tau,k)\ee^{2\ii k^2 (t-\tau)\sigma_3}\ee^{\ii kx\sigma_3} \notag\\
& +\int_0^x \ee^{-\ii k(x-y)\sigma_3}
U(y,t)\Phi_1(y,t,k)\ee^{\ii k(x-y)\sigma_3}\dd y,\label{phi1}\\
\Phi_2(x,t,k)&= E +\ee^{-\ii kx\sigma_3}\int_0^t \ee^{-2\ii k^2 (t-\tau)\sigma_3}
    V(0,\tau,k) \Phi_2(0,\tau,k)\ee^{2\ii k^2 (t-\tau)\sigma_3}\ee^{\ii kx\sigma_3} \notag\\
& +\int_0^x \ee^{-\ii k(x-y)\sigma_3}
U(y,t)\Phi_2(y,t,k)\ee^{\ii k(x-y)\sigma_3}\dd y,\label{phi2}\\
\Phi_3(x,t,k)&=\ee^{\ii(\omega t -\beta x)\sigma_3}{\cal E}(k)-\int_x^{\infty}\Gamma^p(x,y,t,k)
\left[U(y,t)-U^p(y,t)\right]\Phi_3(y,t,k)\ee^{\ii X(k)(x-y)\sigma_3}\dd y, \label{phi3}
\end{align}
\end{subequations}
where $E$ is the $2\times 2$ identity matrix,
with  
$$
\Gamma^p(x,y,t,k) := \Psi^p(x,t,k)\left(\Psi^p(y,t,k)\right)^{-1}.
$$
Notice that $\Gamma^p(x,y,t,k)$ is an entire function, as the solution of the Cauchy
problem $\Gamma^p_x+\ii k \sigma_3 \Gamma^p = U^p \Gamma^p$, $\Gamma^p(y)=E$.

From (\ref{Lax1}) and  (\ref{Lax2}) it follows that 
$\det \Phi_j=\det \Psi_j\equiv 1$ for all $k$ where the corresponding $\Phi_j(k)$ is defined.
From (\ref{phi}) it follows that  $\Phi_1(\cdot,\cdot,k)$ and $\Phi_2(\cdot,\cdot,k)$ are entire in $k$
whereas $\Phi_3(\cdot,\cdot,k)$ is columnwise analytic: the first column 
$\Phi_3^{(1)}(\cdot,\cdot,k)$ of $\Phi_3$ is analytic in ${\mathbb{C}}_-\setminus\gamma$
and the second column 
$\Phi_3^{(2)}(\cdot,\cdot,k)$ is analytic in ${\mathbb{C}}_+\setminus\gamma$.
In what follows, we will denote by $f_\pm(k)$, $k\in\gamma$  the limiting values of a function $f(k)$ as $k$ approaches 
$\gamma$ from
the left $(+)$ or from the right $(-)$.

Besides, the whole matrices $(\Phi_3)_\pm (\cdot,\cdot,k)$ are determined for $k$ on 
$\gamma_\pm$, the corresponding side of $\gamma$.
In order to describe them (or $(\Psi_3)_\pm (\cdot,\cdot,k)$ ), it is convenient to
introduce the $2\times 2$  function $\mu(x,t,k)$, $k\in \gamma$ as  the solution of the integral
equation
$$
\mu(x,t,k)=E -\int_x^{\infty}\Gamma^p(x,y,t,k)
\left[U(y,t)-U^p(y,t)\right]\mu(y,t,k)(\Gamma^p)^{-1}(x,y,t,k)\dd y, 
$$
in terms of which we have 
$$
\Psi_3(x,t,k) = \mu(x,t,k)\Psi^p(x,t,k).
$$
Thus 
\begin{equation}\label{psi3-lim}
 (\Psi_3)_\pm(x,t,k) = \mu(x,t,k)\Psi^p_\pm(x,t,k), \qquad k\in\gamma.
\end{equation}

Now  define the scattering matrices $s(k)$ and $S(k)$, $k\in {\mathbb{R}}\cup\gamma_+\cup\gamma_-$,
 as the matrices relating the eigenfunctions
$\Psi_j(x,t,k)$ for all $x$ and $t$:
\begin{equation}\label{scat}
 \Psi_3(x,t,k)=\Psi_2(x,t,k)s(k),\qquad \Psi_1(x,t,k)=\Psi_2(x,t,k)S(k).
\end{equation}
From (\ref{phi}) and (\ref{scat}) it follows that 
$$
s(k)=\Psi_3(0,0,k).
$$
Thus the the symmetry 
\begin{equation}\label{sym}
 \overline{(\Psi_j)_{11}(x,t,\bar k)}=(\Psi_j)_{22}(x,t,k),\qquad \overline{(\Psi_j)_{12}(x,t,\bar k)}=-(\Psi_j)_{21}(x,t,k)
\end{equation}
implies that 
\begin{equation}\label{s-S}
s(k)=\begin{pmatrix}
      \bar a(k) & b(k) \\ -\bar b(k) & a(k)
     \end{pmatrix},\qquad S(k)=\begin{pmatrix}
      \bar A(k) & B(k) \\ -\bar B(k) & A(k)
     \end{pmatrix}, \qquad k\in \mathbb{R}.
\end{equation}
Moreover, $\det s(k)=\det S(k)=1$.

It follows from the definition of $a(k)$ and $b(k)$ and the analyticity properties of $\Psi_3$ that
the spectral functions $a(k)$ and $b(k)$ are analytic in $k\in {\mathbb C}^+\setminus \gamma$; moreover,
 $a(k)\to 1 $ and $b(k)\to 0$
as $k\to\infty$, and the limiting values $a_\pm(k)$ and $b_\pm(k)$, $k\in \gamma$ are related 
as follows:
\begin{lem}\label{a-b-lim}
\begin{equation}\label{ga1-1}
 a_-(k)b_+(k)-a_+(k)b_-(k)=\ii, \qquad k\in \gamma.
\end{equation}
\end{lem}

Indeed, 
 the representation (\ref{psi3-lim}) yields
$$
s_\pm(k)=\mu(0,0,k){\cal E}_\pm(k).
$$
From (\ref{E0}) and (\ref{x-om-nu}) we have $\nu_+(k) = \ii\nu_-(k)$ and thus 
${\cal E}_+(k)=\ii{\cal E}_-(k)\begin{pmatrix}
                                 0 & 1 \\
                                 1 & 0
                                \end{pmatrix}$, which yields
$$
s_+(k)=\ii s_-(k)\begin{pmatrix}
                                 0 & 1 \\
                                 1 & 0
                                \end{pmatrix}.
$$
In terms of $a_\pm$ and $b\pm$ this reads
$$
s_+(k)=\begin{pmatrix}
                                 \ii b_-(k) & b_+(k) \\
                                 \ii a_-(k) & a_+(k)
                                \end{pmatrix}, 
\qquad 
s_-(k)=\begin{pmatrix}
                                 -\ii b_+(k) & b_-(k) \\
                                 -\ii a_+(k) & a_-(k)
                                \end{pmatrix}.
$$
Finally, the determinant relation $\det s \equiv 1$ gives (\ref{ga1-1}).

\hfill $\Box$

\medskip

In the particular case of pure exponential initial data,
$u(x,0) = \alpha \ee^{-2\ii\beta x}$ for $x>0$,
we have
$s(k) =\Psi_3(0,0,k) =  {\cal E}(k)$
and thus 
\begin{equation}\label{a-b}
\begin{aligned}
a(k) &= \overline{a(\bar k)} = \frac{1}{2}\left(\nu(k)+\dfrac{1}{\nu(k)}\right), \\
b(k) &= -\overline{b(\bar k)} = \frac{1}{2}\left(\nu(k)-\dfrac{1}{\nu(k)}\right).
\end{aligned}
\end{equation}
In this case, the identity (\ref{ga1-1}) is seen directly.

The spectral functions $A=A(k;T)$ and $B=B(k;T)$ defined by (\ref{s-S}) and (\ref{scat}) 
in terms of $\Psi_1(0,0,k)$,
are entire functions bounded in the domains $I$ and $III$,
where $I=\{k: \Im k>0, \Re k >0\}$ and $III=\{k: \Im k<0, \Re k <0\}$,
 with $A(k;T)\to 1 $ and $B(k;T)\to 0$
as $k\to\infty$.
Moreover, they are determined by $u(0,t)$ and $u_x(0,t)$ for  $0\le t\le T$ only,
via 
$$
S(k;T)=(\Psi_2)^{-1}(0,t,k)\Psi_1(0,t,k)
$$
for any $t\in [0,T]$.

The compatibility of the set of functions $\{u(x,0),u(0,t),u_x(0,t)\}$ as traces
of a solution $u(x,t)$ of the NLS equation can be expressed in terms of the associated
spectral functions as follows \cite{FIS05}:
\begin{equation}\label{gr}
 A(k;T)b(k)-a(k)B(k;T) = c(k;T)\ee^{4\ii k^2 T}, \qquad k\in {\mathbb C}^+\setminus \gamma,
\end{equation}
with some $c(k;T)=O(\frac{1}{k})$ as $k\to\infty$
(in the general scheme \cite{F02} of analysis of IBV problems, (\ref{gr}) is called the
\emph{global relation}).

Define 
\begin{equation}\label{d}
 d(k):=a(k)\overline{A(\bar k)}+b(k)\overline{B(\bar k)}, \qquad k\in II\setminus \gamma,
\end{equation}
where $II =\{k:\Im k>0, \Re k <0\}$.
Finally, assuming  that $d(k)$
has at most a finite number of simple zeros in II,
define a piecewise meromorphic function (the superscripts denote the  column of the respective matrix)
$M(x,t,k)$, $k\in {\mathbb C}\setminus\{{\mathbb R}\cup \ii{\mathbb R}\cup\gamma\}$:
\begin{equation}\label{M}
M(x,t,k)=\begin{cases}
           \begin{pmatrix}
            \dfrac{\Psi_2^{(1)}}{a}\ \  \Psi_3^{(2)}\end{pmatrix}
            \ee^{\ii(kx+2k^2 t)\sigma_3}, & k\in I=\{k:\ \Im k>0, \Im k^2>0\}\\
     \begin{pmatrix} \dfrac{\Psi_1^{(1)}}{d}\ \  \Psi_3^{(2)}\end{pmatrix}\ee^{\ii(kx+2k^2 t)\sigma_3}, &
            k\in II=\{k:\ \Im k>0, \Im k^2<0\}\\
  \begin{pmatrix} \Psi_3^{(1)} \ \  \dfrac{\Psi_1^{(2)}}{\bar d}\end{pmatrix}\ee^{\ii(kx+2k^2 t)\sigma_3}, &
            k\in III =\{k:\ \Im k<0, \Im k^2>0\}\\
 \begin{pmatrix} \Psi_3^{(1)} \ \  \dfrac{\Psi_2^{(2)}}{\bar a}\end{pmatrix}\ee^{\ii(kx+2k^2 t)\sigma_3}, &
            k\in IV =\{k:\ \Im k<0, \Im k^2<0\}
          \end{cases}.
\end{equation}
Then $M(\cdot,\cdot,k) = E +O(1/k)$ as $k\to\infty$,  and
the scattering relations (\ref{scat}) imply that the limiting
values of $M$ on ${\mathbb R}\cup \ii{\mathbb R}\cup\gamma$ satisfy the jump relations
\begin{equation}\label{M-jump}
 M_+(x,t,k)=M_-(x,t,k)\ee^{-\ii(kx+2k^2 t)\sigma_3}J_0(k)\ee^{\ii(kx+2k^2 t)\sigma_3},
\end{equation}
where $J_0(k)$ is defined as follows.
\begin{enumerate}
\item
For $k\in {\mathbb R}\cup \ii{\mathbb R}$, 
\begin{equation}\label{J0}
 J_0(k):=\begin{cases}
           \begin{pmatrix}
1+|r(k)|^2 & \bar r(k) \\ r(k) & 1
\end{pmatrix}, & k>0, \\
     \begin{pmatrix} 1 & 0 \\ \Gamma(k) & 1
\end{pmatrix}, &  k\in \ii\mathbb{R}_+,\\
  \begin{pmatrix} 1 & \bar \Gamma(\bar k) \\ 0 & 1 \end{pmatrix}, &
            k\in \ii\mathbb{R}_-,\\
 \begin{pmatrix} 1+|r(k)+\Gamma(k)|^2  & \bar r(k) + \bar \Gamma(k) \\
  r(k) + \Gamma(k) & 1
 \end{pmatrix}, & k<0,
          \end{cases}
\end{equation}
where 
\begin{equation}\label{ga}
 r(k):=\frac{\bar b(k)}{a(k)}, \qquad \Gamma(k) := -\dfrac{\bar B(\bar k)}{a(k)d(k)}
\equiv - \dfrac{\bar B(\bar k)/\bar A(\bar k)}{a(k)\left(a(k)+b(k)\bar B(\bar k)/\bar A(\bar k)\right)}. 
\end{equation}
Here the orientation of the contour is chosen as from $-\infty$ to $+\infty$
along $\mathbb{R}$ and away from $0$ along $\ii\mathbb{R}$.
\item
For  $k\in \gamma$, with the orientation of $\gamma$  upward,
the definition of $J_0(k)$ 
depends on the sign of $\beta$:
\begin{itemize}
\item
if $\beta>0$, then $\gamma\in I\cup IV$, and 
\begin{equation}\label{j0+prelim}
J_0(k) := \begin{cases}
\begin{pmatrix} \frac{a_-(k)}{a_+(k)} & \Gamma_1(k)\\
0 & \frac{a_+(k)}{a_-(k)}
\end{pmatrix}, & k\in \gamma\cap I,\\
\begin{pmatrix} \frac{\bar a_+(\bar k)}{\bar a_-(\bar k)} & 0 \\
-\bar \Gamma_1(\bar k) & \frac{\bar a_-(\bar k)}{\bar a_+(\bar k)}
\end{pmatrix}, & k\in \gamma\cap IV
\end{cases}
\end{equation}
with 
\begin{equation}\label{Ga1}
\Gamma_1(k) := b_+(k)a_-(k) - b_-(k)a_+(k).
\end{equation}
Taking into account (\ref{ga1-1}), we have $\Gamma_1(k)\equiv\ii$ and thus
the jump (\ref{psi3-lim}) takes the form
\begin{equation}\label{j0+}
J_0(k) = \begin{cases}
\begin{pmatrix} \frac{a_-(k)}{a_+(k)} & \ii\\
0 & \frac{a_+(k)}{a_-(k)}
\end{pmatrix}, & k\in \gamma\cap I,\\
\begin{pmatrix} \frac{\bar a_+(\bar k)}{\bar a_-(\bar k)} & 0 \\
\ii & \frac{\bar a_-(\bar k)}{\bar a_+(\bar k)}
\end{pmatrix}, & k\in \gamma\cap IV
\end{cases}
\end{equation}

\item
if $\beta<0$, then $\gamma\in II\cup III$, and 
\begin{equation}\label{j0-prelim}
J_0(k) := \begin{cases}
\begin{pmatrix} \frac{d_-(k)}{d_+(k)} & \hat\Gamma_1(k)\\
0 & \frac{d_+(k)}{d_-(k)}
\end{pmatrix}, & k\in \gamma\cap II,\\
\begin{pmatrix} \frac{\bar d_+(\bar k)}{\bar d_-(\bar k)} & 0 \\
-\bar{\hat\Gamma}_1(\bar k) & \frac{\bar d_-(\bar k)}{\bar d_+(\bar k)}
\end{pmatrix}, & k\in \gamma\cap III
\end{cases}
\end{equation}
with $\hat\Gamma_1(k) := (A(k)b_+(k)-B(k)a_+(k))d_-(k)-(A(k)b_-(k)-B(k)a_-(k))d_+(k)$.
\end{itemize}
Direct calculations, taking into account the determinant equality $A(k)\bar A(\bar k)+B(k)\bar B(\bar k)=1$
(coming from $\det S\equiv 1$)
show that $\hat\Gamma_1(k)=\Gamma_1(k)$ and thus $\hat\Gamma_1(k)\equiv \ii$, which yields
\begin{equation}\label{j0-}
J_0(k) = \begin{cases}
\begin{pmatrix} \frac{d_-(k)}{d_+(k)} & \ii \\
0 & \frac{d_+(k)}{d_-(k)}
\end{pmatrix}, & k\in \gamma\cap II,\\
\begin{pmatrix} \frac{\bar d_+(\bar k)}{\bar d_-(\bar k)} & 0 \\
\ii  & \frac{\bar d_-(\bar k)}{\bar d_+(\bar k)}
\end{pmatrix}, & k\in \gamma\cap III.
\end{cases}
\end{equation}
\end{enumerate}
Complemented with the normalization condition $M=E+O(1/k)$ as $k\to\infty$
and the respective residue conditions at the zeros of  $d(k)$ and $a(k)$
(see \cite{FIS05} for details),
the jump relation (\ref{M-jump}) can be viewed as the \emph{Riemann--Hilbert problem}:

given $\{a(k),b(k),A(k),B(k)\}$ (and the residue parameters, if any), find $M(x,t,k)$ for all $x\ge 0$ and $t\ge 0$.
Then the solution of the NLS equation, $u(x,t)$, is given in terms of $M(x,t,k)$
by 
\begin{equation}\label{u-RHP}
 u(x,t)=2\ii \lim_{k\to\infty}k M_{12}(x,t,k).
\end{equation}
Moreover, $u(x,0)$ generates $\{a(k),b(k)\}$  and 
$\{u(0,t),u_x(0,t)\}$ generates $\{A(k),B(k)\}$ as the corresponding 
spectral functions provided the latter verify the global relation (\ref{gr}).
Therefore, the Riemann--Hilbert problem approach gives the solution
of the overdetermined IBV problem
\begin{equation} \label{ibv-g}
\begin{aligned}
  & \ii u_t + u_{xx} + 2|u|^2 u =0, \qquad & x>0, t>0,\\
& u(x,0)=u_0(x),& x\ge 0, \\
& u(0,t)=v_0(t), & 0\le t\le T,  \\
& u_x(0,t)=v_1(t), &0\le t\le T
\end{aligned}
\end{equation}
provided that the spectral functions $\{a(k),b(k),A(k),B(k)\}$
constructed from $\{u_0(x),v_0(t),v_1(t)\}$ satisfy the global relation (\ref{gr}).

\section{The Riemann--Hilbert formalism for Robin boundary condition}
\setcounter{equation}{0}

The overdetermined nature of the Riemann--Hilbert problem approach
to the IBV problems can be overcome in certain special cases of boundary conditions
\cite{F02}. An efficient way to do this is the use of an additional symmetry in the spectral problem for the $t$-equation
of the Lax pair. 

Since the $t$-equation, considered alone, is independent of the initial conditions,
the following symmetry considerations, associated with the Robin boundary condition, are 
exactly the same as in \cite{FIS05, IS12}, where the IBV with decaying initial conditions were considered.
Namely,  if $u+qu_x=0$ with some $q\in \mathbb{R}$, then the matrix 
$\tilde V:=V-2\ii k^2\sigma_3$ of the $t$-equation $\Psi_t=\tilde V \Psi$
satisfies the symmetry relation \cite{FIS05}
\begin{equation}\label{V-sym}
 \tilde V(x,t,-k)=N(k)\tilde V(x,t,k) N^{-1}(k),
\end{equation}
where $N(k)=\diag\{N_1(k), N_2(k)\}$ with $N_1(k)= 2k+\ii q$ and $N_2(k)= -2k+\ii q$.
In turn, (\ref{V-sym}) implies the symmetry for $S$: $S(-k;T)=N(k)S(k;t)N^{-1}(k)$,
which reads in terms of $A$ and $B$ as 
\begin{eqnarray}\label{AB-sym}
 A(-k;T)&=&A(k;T), \nonumber\\
B(-k;T)&=&-\frac{2k+\ii q}{2k-\ii q}B(k;T).
\end{eqnarray}

Now we observe that the global relation (\ref{gr}) also has the same form
as in the case of decaying initial conditions, which, being combined with 
 the symmetry 
relation (\ref{AB-sym}), 
suggests
 rewriting the RH problem (\ref{M-jump})
in the form that uses only the spectral functions $a(k)$ and $b(k)$
associated with the initial values $u(x,0)$. Namely, 
 since the exponential in the r.h.s.
of (\ref{gr}) is rapidly decaying for  $k\in I$,
the global relation (\ref{gr}) suggests to replace 
$\frac{B}{A}(k;T)$ by $\frac{b}{a}(k)$ for $k\in I$. Then, the symmetry 
relation (\ref{AB-sym}) suggests to replace 
\begin{equation}\label{repl1}
\frac{B}{A}(k;T) \mapsto -\frac{2k-\ii q}{2k+\ii q}\frac{b}{a}(-k) \qquad \text{ for}\   k\in III
\end{equation}
 and consequently
to replace 
\begin{equation}\label{repl2}
\frac{\bar B}{\bar A}(\bar k;T)\mapsto  
-\frac{2k+\ii q}{2k-\ii q}\frac{\bar b}{\bar a}(-\bar k)\qquad \text{ for}\  k\in II.
\end{equation}

\begin{rem}
We emphasize that we are not going to compare the RH problem obtained as a result of
the replacement described above, with the original RH problem (\ref{M-jump})--(\ref{j0-}).
Instead, we will show directly that the resulting RH problem (for $\hat M$, see below)
gives the solution of the IBV problem (\ref{ibv}).
\end{rem}
\begin{rem}
A principal difference when comparing with the case of decaying initial conditions
is that now $a(k)$ and $b(k)$ (as well as their ratio) has jump discontinuities across $\gamma$,
which affects the jump conditions on $\gamma$ and eventually on $\gamma_1$, where $\gamma_1$ is symmetric 
to  $\gamma$ with respect to the imaginary axis:
$$
\gamma_1 := \{k=k_1+\ii k_2: k_1=-\beta, k_2\in(-\alpha,\alpha)\}
$$
(the latter is due to the jump of $\bar a(-\bar k)$ and $\bar b(-\bar k)$ across $\gamma_1$).
\end{rem}

The resulting jump conditions 
\begin{equation}\label{M-t-jump}
 \tilde M_+(x,t,k)=\tilde M_-(x,t,k)\ee^{-\ii(kx+2k^2 t)\sigma_3}\tilde J_0(k)\ee^{\ii(kx+2k^2 t)\sigma_3},
\end{equation}
are as follows.
\begin{enumerate}
\item
For $k\in \mathbb{R}\cup \ii\mathbb{R}$,  $\tilde J_0(k)$ has the same form as 
 in (\ref{J0}),
but with $\Gamma(k)$ replaced, accordingly to (\ref{ga}) and (\ref{repl2}),
 by  $\tilde\Gamma(k)$ (cf. \cite{FIS05,IS12}), where
\begin{equation}\label{ga-t}
 \tilde\Gamma(k) := \frac{\overline{b(-\bar k)}}{a(k)}
\frac{2k+\ii q}{(2k-\ii q)a(k)\overline{a(-\bar k)}-(2k+\ii q)b(k)\overline{b(-\bar k)}}:
\end{equation}

\begin{equation}\label{j-t-R}
 \tilde J_0(k):=\begin{cases}
           \begin{pmatrix}
1+|r(k)|^2 & \bar r(k) \\ r(k) & 1
\end{pmatrix}, & k>0, \\
     \begin{pmatrix} 1 & 0 \\ \tilde\Gamma(k) & 1
\end{pmatrix}, &  k\in \ii\mathbb{R}_+,\\
  \begin{pmatrix} 1 & \bar{\tilde\Gamma}(\bar k) \\ 0 & 1 \end{pmatrix}, &
            k\in \ii\mathbb{R}_-,\\
 \begin{pmatrix}
1+|r_e(k)|^2 & \bar r_e(k) \\
r_e(k) & 1 
\end{pmatrix}, & k<0,
          \end{cases}
\end{equation}
and
\begin{equation}\label{r-e}
 r_e(k):=r(k)+\tilde \Gamma(k) = \frac{(2k-\ii q )\overline{b(k)}\overline{a(-k)}+
(2k+\ii q) \overline{b(-k)}\overline{a(k)}}{(2k-\ii q) a(k)\overline{a(-k)}-
(2k+\ii q) \overline{b(-k)}b(k)}.
\end{equation}
\item
For $k\in\gamma$, the jump for $\tilde M$ depends on the sign of $\beta$.
\begin{itemize}
\item
if $\beta>0$, then $\tilde J_0(k)=J_0(k)$, see (\ref{j0+}) (notice that in this case, $J_0(k)$ is already defined in terms of 
$a(k)$ and $b(k)$ only).
\item
if $\beta<0$, then $\gamma\in II\cup III$, and
$$
\frac{d_-(k)}{d_+(k)} \equiv \frac{a_-(k) + b_-(k)\frac{\bar B(\bar k)}{\bar A(\bar k)}}
{a_+(k) + b_+(k)\frac{\bar B(\bar k)}{\bar A(\bar k)}}
$$
  is replaced, accordingly to (\ref{repl2}), by 
\begin{equation}\label{d-d}
\delta(k):=\frac{a_-(k) - b_-(k)\frac{2k+\ii q}{2k-\ii q}\frac{\bar b(-\bar k)}{\bar a(-\bar k)}}
{a_+(k) - b_+(k)\frac{2k+\ii q}{2k-\ii q}\frac{\bar b(-\bar k)}{\bar a(-\bar k)}} = 
\frac{(2k-\ii q)a_-(k)\overline{a(-\bar k)} - (2k+\ii q)b_-(k)\overline{b(-\bar k)}}
{(2k-\ii q)a_+(k)\overline{a(-\bar k)} - (2k+\ii q)b_+(k)\overline{b(-\bar k)}},
\end{equation}
which leads to 
 \begin{equation}\label{J0-ga}
\tilde J_0(k) = 
\begin{cases}
\begin{pmatrix} \delta(k) & \ii \\
0 & \delta^{-1}(k)
\end{pmatrix}, & k\in \gamma\cap{\mathbb C}_+,\\
\begin{pmatrix} \bar\delta^{-1}(\bar k) & 0 \\
\ii & \bar\delta(\bar k)
\end{pmatrix}, & k\in \gamma\cap {\mathbb C}_-
\end{cases}
\end{equation} 
\end{itemize}
\item
In the case $\beta>0$, the fact that $\tilde \Gamma(k)$ has an {\em additional} (with respect to $\Gamma(k)$) jump
across $\gamma_1$
suggests {\em introducing} the jump conditions for $\tilde M$ on $\gamma_1$:
\begin{equation}\label{J0-ga1}
\tilde J_0(k) = \begin{cases}
\begin{pmatrix}
1 & 0 \\
\tilde\Gamma_+(k) - \tilde\Gamma_-(k) & 1 
\end{pmatrix}, & k\in \gamma_1\cap {\mathbb C}_+ \\
\begin{pmatrix}
1 & \bar{\tilde\Gamma}_-(k) - \bar{\tilde\Gamma}_+(k) \\
0 & 1 
\end{pmatrix}, & k\in \gamma_1\cap {\mathbb C}_-
\end{cases}
\end{equation}
(actually, in this case $\gamma_1 \in II\cup III$ whereas $\gamma \in I\cup IV$).
\end{enumerate}

At this stage, the jump conditions in the cases $\beta>0$ and $\beta<0$ look quite different.
But now
we notice that the exponentials 
in $\begin{pmatrix}
     1 & 0 \\ \tilde \Gamma(k) \ee^{2\ii k x + 4\ii k^2 t} & 1
    \end{pmatrix}
$
and 
$\begin{pmatrix}
     1 & \overline{\tilde \Gamma(\bar k)} \ee^{-2\ii k x - 4\ii k^2 t} \\ 0 & 1
    \end{pmatrix}
$
are bounded in respectively I and IV.  
Therefore, we can deform the RH problem with jump 
(\ref{M-t-jump}) on $\mathbb{R}\cup \ii\mathbb{R}\cup\gamma\cup\gamma_1$ to that on 
$\mathbb{R}\cup\gamma\cup\gamma_1$ (thus getting rid of the jump on $\ii\mathbb{R}$)
 by introducing
\begin{equation}\label{M-h}
 \hat M(x,t,k)=\begin{cases}
               \tilde  M(x,t,k), & k\in II\cup III, \\
\tilde M(x,t,k)\begin{pmatrix}
     1 & 0 \\ \tilde \Gamma(k) \ee^{2\ii k x + 4\ii k^2 t} & 1
    \end{pmatrix}, & k\in I, \\
\tilde M(x,t,k)\begin{pmatrix}
     1 & -\overline{\tilde \Gamma(\bar k)} \ee^{-2\ii k x - 4\ii k^2 t} & 1 \\ 0 & 1
    \end{pmatrix}, & k\in IV.
               \end{cases}
\end{equation}
The resulting jump conditions take the form
\begin{equation}\label{M-h-jump}
 \hat M_+(x,t,k)=\hat M_-(x,t,k)\ee^{-\ii(kx+2k^2 t)\sigma_3}\hat J_0(k)\ee^{\ii(kx+2k^2 t)\sigma_3},
\quad k\in \mathbb{R}\cup\gamma\cup\gamma_1,
\end{equation}
where:
\begin{enumerate}
\item
for $k\in\mathbb{R}$,
\begin{equation}\label{j-0-h}
 \hat J_0(k) = \begin{pmatrix}
                1+|r_e(k)|^2 & \bar r_e(k) \\ r_e(k) & 1
               \end{pmatrix}
\end{equation}
with $r_e(k)$ defined by (\ref{r-e}).
\item
\begin{itemize}
\item 
in the case $\beta>0$, direct calculations give
$$
 \hat J_0(k) = \begin{pmatrix}
 1 & 0 \\
 -\tilde\Gamma_-(k) & 1
 \end{pmatrix}
 \begin{pmatrix}
 \frac{a_-(k)}{a_+(k)} & \ii \\
0 & \frac{a_+(k)}{a_-(k)}
 \end{pmatrix}
 \begin{pmatrix}
 1 & 0 \\
 \tilde\Gamma_+(k) & 1
 \end{pmatrix} = \begin{pmatrix}
 \delta(k) & \ii \\
0 & \delta^{-1}(k)
 \end{pmatrix}\quad \text{for}\ k\in\gamma\cap I
$$
and 
$$
 \hat J_0(k) = \begin{pmatrix}
 1 & \bar{\tilde\Gamma}_-(\bar k) \\
  0 & 1
 \end{pmatrix}
 \begin{pmatrix}
 \frac{\bar a_+(\bar k)}{\bar a_-(\bar k)} & 0 \\
\ii &  \frac{\bar a_-(\bar k)}{\bar a_+(\bar k)}
 \end{pmatrix}
 \begin{pmatrix}
 1 & -\bar{\tilde\Gamma}_+(\bar k) \\
0 & 1
 \end{pmatrix} = \begin{pmatrix}
 \bar\delta^{-1}(\bar k) & 0 \\
\ii & \bar\delta(\bar k)
 \end{pmatrix}\quad \text{for}\ k\in\gamma\cap IV
$$
and thus the jump takes the form of (\ref{J0-ga}) as in the case $\beta<0$.
On the other hand, the jump is not changed for $k\in \gamma_1$: $ \hat J_0(k) =  \tilde J_0(k)$ as in (\ref{J0-ga1}).
\item
 in the case $\beta<0$, a new jump occurs on $\gamma_1$, due to the jump of $\tilde\Gamma$, having the form (\ref{J0-ga1})
 as in the case $\beta>0$:
 \begin{equation}\label{J0-ga1-}
\tilde J_0(k) = \begin{cases}
\begin{pmatrix}
1 & 0 \\
\tilde\Gamma_+(k) - \tilde\Gamma_-(k) & 1 
\end{pmatrix}, & k\in \gamma_1\cap I \\
\begin{pmatrix}
1 & \bar{\tilde\Gamma}_-(k) - \bar{\tilde\Gamma}_+(k) \\
0 & 1 
\end{pmatrix}, & k\in \gamma_1\cap IV
\end{cases}.
\end{equation}
On the other hand, 
the jump is not changed for $k\in \gamma$: $ \hat J_0(k) =  \tilde J_0(k)$ as in (\ref{J0-ga}).
\end{itemize}
\end{enumerate}

Summarizing, in the both cases, $\beta>0$ and $\beta<0$, we arrive at 

\begin{prop}\label{rep}
Assume that $u(x,t)$ is the solution of the IBV problem (\ref{ibv}).
Then 
\begin{equation}\label{u-h}
u(x,t)=2\ii\lim_{k\to\infty}k\hat M_{12}(x,t,k),
\end{equation}
where $\hat M$ is a piece-wise meromorphic function satisfying 
jump conditions (\ref{M-h-jump}).
The jump matrix is given by (\ref{j-0-h}) for $k\in \mathbb{R}$ and by 
$\hat J(k)=\tilde J(k)$ for $k\in\gamma\cup\gamma_1$, where $\tilde J(k)$ is defined 
in terms of $a(k)$, $b(k)$ and $q$
by
(\ref{J0-ga}) and (\ref{J0-ga1}) taking into account (\ref{ga-t}) and (\ref{d-d}).
Here $a(k)$ and $b(k)$ are determined by the initial values $u(x,0)=u_0(x)$, $x>0$
through 
$$
\begin{pmatrix}
b(k)\\a(k)
\end{pmatrix} = \Psi_3^{(2)} (0,0,k) = \Phi_3^{(2)} (0,0,k),
$$
where $\Phi_3^{(2)} (x,0,k)$ is the solution of (\ref{phi3}) with $U(y)=\begin{pmatrix}
0 & u_0(y)\\ \bar u_0(y) & 0
\end{pmatrix}$, see (\ref{U}).
\end{prop}

If the denominator in (\ref{r-e}) has no zeros in  ${\mathbb{C}}^+$, then 
$\hat M$ is uniquely determined as the solution of 
{\em the Riemann--Hilbert (RH) problem}: given  $a(k)$, $b(k)$ and $q$, 
find a piecewise analytic (relative to ${\mathbb{R}\cup\gamma
\cup\gamma_1}$),
matrix-valued function  satisfying the jump conditions from Proposition \ref{rep}
and the normalization condition
 $\hat M\to E $ as $k\to\infty$. 
 In case the denominator in (\ref{r-e}) has  zeros, the formulation of the RH problem 
is to be complemented by the 
residue conditions at these points. Introducing $a_e(k)$ by (\ref{a-e}), see below,
and assuming that the zeros $k_j$, $j=1,\dots,N$ of $a_e(k)$ 
($\Im k_j>0$)
are simple and do not coincide with possible zeros of $a(k)$, the residue conditions are 
as follows (cf. \cite{FIS05}):
\begin{eqnarray}\label{res}
\R{Res}_{k_j}\hat M^{(1)}(x,t,k) &=& 
\frac{\overline{b(-\bar k_j)}\ee^{2\ii\Theta(x,t,k_j)}}{\dot a_e(k_j)a(k_j)}\chi(k_j)\hat M^{(2)}(x,t,k_j),\nonumber\\
\R{Res}_{\bar k_j}\hat M^{(2)}(x,t,k) &=&
\frac{b(-\bar k_j)\ee^{-2\ii\bar\Theta(x,t,k_j)}}{\overline{\dot a_e(k_j)}\overline{a(k_j)}}\bar \chi(k_j)\hat M^{(1)}(x,t,\bar k_j),
\end{eqnarray}
where $\Theta(x,t,k)=kx+2k^2 t$ and 
\begin{equation}\label{chi}
 \chi(k) = \begin{cases}
\frac{2k+\ii q}{2k-\ii q}, & \text{case 1}\\
1, & \text{case 2}
\end{cases}
\end{equation}
(for the definition of cases 1 and 2, see (\ref{a-e}) and (\ref{b-e}) below).

Comparing this with the case of the Robin IBV problem with decaying initial data \cite{IS12}, the RH problem
in the oscillatory case has two additional jumps, across $\gamma$ and $\gamma_1$.
On the other hand, it has the same structure as in the case of the shock IV problem \cite{BV07},
which of course is not surprising because, as we discussed in Introduction, the shock IV problem
in \cite{BV07} corresponds to $q=0$ in (\ref{ibv}).

Now, having the RH problem formulated in terms of the spectral functions $a(k)$ and $b(k)$ associated
with the initial data in (\ref{ibv}) and of the Robin constant $q$, we can prove directly (without making reference to 
the boundary values and the spectral functions  $A(k)$ and $B(k)$ associated with them)
the following Theorem (we formulate it for the case of pure exponential initial data in (\ref{ibv});
for the general case, see Remark \ref{ini-gen} below):
\begin{thm}\label{main-thm}
Let $\hat M(x,t,k)$ be the solution of the Riemann--Hilbert problem with the jump conditions (\ref{M-h-jump}) and  residue conditions (\ref{res}) determined
in terms $q$ and the spectral functions $a(k)$ and $b(k)$ (\ref{a-b}) via (\ref{r-e})--(\ref{j-0-h}).
Let $u(x,t)$ be determined in terms of $\hat M(x,t,k)$ via (\ref{u-h}).
Then $u(x,t)$ is the solution of the IBV problem (\ref{ibv}) with $u_0(x)=\alpha\ee^{-2\ii\beta x}$.
\end{thm}

\noindent Proof.

(i) The function determined by (\ref{u-h}) via the solution $\hat M$ of the  RH problem 
satisfies the
NLS equation in the domain $x>0$, $t>0$; this is a standard fact
based on ideas of the dressing method, see, e.g., \cite{FT}. 

(ii) In order to verify the initial condition $u(x,0)=u_0(x)$, one observes that 
for $t=0$ and $x>0$, the exponentials 
in $\begin{pmatrix}
     1 & 0 \\ \tilde \Gamma(k) \ee^{2\ii k x} & 1
    \end{pmatrix}
$
and 
$\begin{pmatrix}
     1 & \overline{\tilde \Gamma(\bar k)} \ee^{-2\ii k x} \\ 0 & 1
    \end{pmatrix}
$
are  bounded in respectively ${\mathbb C}_+$ and ${\mathbb C}_-$. Thus we can deform the RH problem with jump 
(\ref{M-h-jump}) (for $t=0$) as follows: define $\check M(x,0,k)$ by 
\begin{equation}\label{M-c}
 \check M(x,0,k)=\begin{cases}
\hat M(x,0,k)\begin{pmatrix}
     1 & 0 \\ -\tilde \Gamma \ee^{2\ii k x } & 1
    \end{pmatrix}, & k\in {\mathbb C}_+, \\
\hat M(x,0,k)\begin{pmatrix}
     1 & \overline{\tilde \Gamma} \ee^{-2\ii k x } & 1 \\ 0 & 1
    \end{pmatrix}, & k\in {\mathbb C}_-.
               \end{cases}
\end{equation}
This makes (i) the jump across $\gamma_1$ to disappear:
$\check M_+(x,0,k)=\check M_-(x,0,k)$, $k\in \gamma_1$;
(ii) the residue conditions to disappear as well  (since in the case of (\ref{a-b}),
$a(k)$ has no zeros).
On the other hand, the jump across $\mathbb R$ and $\gamma$
take the form

\begin{equation}\label{M-c-jump}
 \check M_+(x,0,k)=\check M_-(x,0,k)\ee^{-\ii kx\sigma_3}\check J_0(k)
\ee^{\ii kx\sigma_3},
\quad k\in \mathbb{R}\cup\gamma,
\end{equation}
where 
\begin{equation}\label{j-0-c}
 \check J_0(k) = \begin{cases}
 \begin{pmatrix}
                1+|r(k)|^2 & \bar r(k) \\ r(k) & 1
               \end{pmatrix}, & k\in \mathbb{R}, \\
               \begin{pmatrix} \frac{a_-(k)}{a_+(k)} & \Gamma_1(k)\\
0 & \frac{a_+(k)}{a_-(k)}
\end{pmatrix}, & k\in \gamma\cap {\mathbb C}_+,\\
\begin{pmatrix} \frac{\bar a_+(\bar k)}{\bar a_-(\bar k)} & 0 \\
-\bar \Gamma_1(\bar k) & \frac{\bar a_-(\bar k)}{\bar a_+(\bar k)}
\end{pmatrix}, & k\in \gamma\cap {\mathbb C}_-
           \end{cases}.
\end{equation}

Finally, introducing 
$$
M_{ini}(x,k):=\begin{cases}
\check M(x,0,k) \begin{pmatrix}
a(k) & 0 \\
0 & \frac{1}{a(k)}
\end{pmatrix}, & \Im k >0 \\
\check M(x,0,k) \begin{pmatrix}
\frac{1}{\bar a(\bar k)}    & 0 \\
0 & \bar a(\bar k)
\end{pmatrix}, & \Im k <0
\end{cases}
$$
the resulting jump conditions for $M_{ini}(x,k)$ are 
\begin{equation}\label{ini-jump}
M_{ini,+}(x,k) = M_{ini,-}(x,k)\ee^{-\ii kx\sigma_3}J_{ini,0}(k)
\ee^{\ii kx\sigma_3},
\quad k\in \mathbb{R}\cup\gamma,
 \end{equation}
where 
\begin{equation}\label{ini}
 J_{ini,0}(k) = \begin{cases}
 \begin{pmatrix}
                1 & \tilde r(k)  \\ 
                 \bar{\tilde r}(k) & 1+|\tilde r(k)|^2 
               \end{pmatrix}, & k\in \mathbb{R}, \\
               \begin{pmatrix} 1 & f(k)\\
0 & 1
\end{pmatrix}, & k\in \gamma\cap {\mathbb C}_+,\\
\begin{pmatrix} 1 & 0 \\
-\bar f(\bar k) & 1
\end{pmatrix}, & k\in \gamma\cap {\mathbb C}_-
           \end{cases},
           \end{equation}
with
$$
\tilde r(k):=\frac{b(k)}{a(k)}
$$
and
\begin{equation}\label{f-r}
f(k):=\tilde r_+(k)- \tilde r_-(k).
\end{equation}

Now we notice that these conditions
coincide with those  for the spectral mapping $\{u_0(x)\} \mapsto \{a(k),b(k)\}$
in the IV problem on the whole line with step-like initial data:
$u_0(x)=\alpha\ee^{-2\ii\beta x}$ for $x>0$ and $u_0(x)=0$ for $x<0$ (see \cite{BKS11}).
This yields $u(x,0)=u_0(x)=\alpha\ee^{-2\ii\beta x}$ for $x>0$ due to the uniqueness 
of the solution of the RH problem.

\begin{rem}\label{ini-gen}
In the case of the general initial data $u_0(x)$ (such that $u_0(x)\to \alpha\ee^{-2\ii\beta x}$ as 
$x\to+\infty$ fast enough), the transformation of the original RH problem to a RH problem
associated with the initial conditions holds as well, the  difference being that in the general case, (i)
the spectral function $f(k)$ is no longer related to $\tilde r(k)$ via (\ref{f-r}) but
is an independent part of the spectral data associated with $u_0(x)$;
(ii) the residue conditions are moved to the location of possible zeros of $a(k)$.
\end{rem}

(iii)
The strategy in proving that $u(x,t)$ satisfies the Robin boundary condition follows 
that in the case of decaying initial data \cite{IS12}. 

First, we notice that  the RH problem (\ref{M-h-jump}) with the jump matrices defined by
(\ref{J0-ga}) for $k\in \gamma$, by (\ref{J0-ga1}) for $k\in \gamma_1$, 
and by (\ref{j-0-h}) for $k\in \mathbb R$ can be considered not only for $x\ge0$ but also 
for $x<0$ thus giving a continuous (at $x=0$, for all $t>0$) continuation of $u(x,t)$
(\ref{u-h}) for $x<0$. The associated continuation of the initial data $u(x,0)$
for $x<0$ will be discussed below; but here we observe that the structure itself
of the jump matrix on $k\in \mathbb R$ suggests {\em interpreting} $r_e(k)$ as a reflection
coefficient for the  problem obtained on the whole line:
$$
r_e(k)=\frac{\bar b_e(k)}{a_e(k)}.
$$
In order to determine $b_e(k)$ and $a_e(k)$ correctly, we require that $a_e(k)$ is analytic 
in ${\mathbb C}\setminus(\gamma\cup\gamma_1)$, $a_e(k)\to 1 $ as $k\to \infty$, 
$|a_e|^{-2}(k) = 1+ |r_e|^2(k)$ for $k\in\mathbb R$, and 
$a_e(k)$ has neither a zero nor a pole at $k=\ii |q|/2$ (cf. \cite{DP10, IS12}).
In view of the definition (\ref{r-e}) of $r_e(k)$, this gives
\begin{equation}\label{a-e}
 a_e(k):=\begin{cases}
         a(k)\overline{a(-\bar k)} - \frac{2k+\ii q}{2k-\ii q}b(k)\overline{b(-\bar k)}, &
	\text{if }\  q<0, a(-\frac{\ii q}{2})\ne 0 \ \text{or }\ q>0, b(\frac{\ii q}{2})=0 \\
 \frac{2k-\ii q}{2k+\ii q}a(k)\overline{a(-\bar k)} 
	- b(k)\overline{b(-\bar k)}, &
	\text{if }\  q>0, b(\frac{\ii q}{2})\ne 0 \ \text{or }\ q<0, a(-\frac{\ii q}{2})=0
        \end{cases}
\end{equation}
and respectively 
\begin{equation}\label{b-e}
 b_e(k):=\begin{cases}
b(k)a(-k) + \frac{2k-\ii q}{2k+\ii q}b(-k)a(k), &
	\text{if }\  q<0, a(-\frac{\ii q}{2})\ne 0 \ \text{or }\ q>0, b(\frac{\ii q}{2})=0 \\
\frac{2k+\ii q}{2k-\ii q}b(k)a(-k) + b(-k)a(k), &
	\text{if }\  q>0, b(\frac{\ii q}{2})\ne 0 \ \text{or }\ q<0, a(-\frac{\ii q}{2})=0
        \end{cases}
\end{equation}
In what follows, we will refer to these two cases of definition of  $a_e(k)$ and $b_e(k)$
as Case 1 and Case 2.

Consequently, from (\ref{d-d}) we have 
$$
\delta(k)=\dfrac{a_{e-}(k)}{a_{e+}(k)},\quad k\in\gamma.
$$

Accordingly to the cases singled out above, let us introduce 
\begin{equation}\label{be}
 h:=\begin{cases}
         \frac{q}{2}, &
	\text{if }\  q<0, a(-\frac{\ii q}{2})\ne 0 \ \text{or }\ q>0, b(\frac{\ii q}{2})=0 \\
-\frac{q}{2}, &
	\text{if }\  q>0, b(\frac{\ii q}{2})\ne 0 \ \text{or }\ q<0, a(-\frac{\ii q}{2})=0
        \end{cases}
\end{equation}
Denote  by $\hat J(x,t,k)$ the jump matrix in (\ref{M-h-jump}), i.e.,
$$
\hat J(x,t,k):=\ee^{-\ii(kx+2k^2 t)\sigma_3}\hat J_0(k)\ee^{\ii(kx+2k^2 t)\sigma_3}.
$$

\begin{lem}
The jump matrix $\hat J_0$ in (\ref{M-h-jump}) satisfies the symmetry conditions:
\begin{eqnarray}\label{J-sym}
\overline{\hat J(0,t,-k)} &=& C_-(k) \hat J(0,t,k) C_+(k)  \qquad \text{for}\ k\in \mathbb R, \\
\left(\overline{\hat J(0,t,-\bar k)}\right)^{-1} &=& C_-(k) \hat J(0,t,k) C_+(k)  \qquad \text{for}\ 
k\in\gamma\cup\gamma_1, \label{sym-g}
\end{eqnarray}
where
\begin{equation}\label{C}
C(k)=\begin{cases}
      \begin{pmatrix}
       a_e(k) & 0 \\ 0 & \frac{1}{a_e(k)}
      \end{pmatrix}
\begin{pmatrix}
 k-\ii h & 0 \\ 0 & k+\ii h
\end{pmatrix}
\sigma_1, & \Im k >0,\\
\begin{pmatrix}
       \frac{1}{\overline{a_e(\bar k)}} & 0 \\ 0 & \overline{a_e(\bar k)}
      \end{pmatrix}
\begin{pmatrix}
 k-\ii{h} & 0 \\ 0 & k+\ii{h}
\end{pmatrix}
\sigma_1, & \Im k < 0,
\end{cases}
\end{equation}
with $\sigma_1 = \begin{pmatrix}
                   0 & 1 \\ 1 & 0
                  \end{pmatrix}
$.
\end{lem}

\noindent Proof. 
(i) The symmetry (\ref{J-sym}) follows from the symmetry of the reflection coefficient:
\begin{equation}\label{r-e-sym1}
 r_e(-k) = r_e(k)\frac{a_e(k)}{\overline{a_e(k)}}\frac{k-\ii{h}}{k+\ii{h}}.
\end{equation}

(ii) The symmetry (\ref{sym-g}) can be checked directly.

\bigskip

\begin{lem}\label{M-sym-lem}
The solution $\hat M$ of the RH problem (\ref{M-h-jump})
satisfies the symmetry condition
\begin{equation}\label{M-sym}
 \overline{\hat M(0,t,-\bar k)} = \sigma_1 \bar P(t) \begin{pmatrix}
                              \frac{1}{k-\ii{h}} & 0 \\ 0 &  \frac{1}{k+\ii{h}}
                             \end{pmatrix}
\bar P^{-1}(t)\hat M(0,t,k)\begin{pmatrix}
 k-\ii{h} & 0 \\ 0 & k+\ii{h}
\end{pmatrix} D(k) \sigma_1,
\end{equation}
where
\begin{equation}\label{D}
D(k)\equiv\diag\{d_1(k), d_2(k)\}=\begin{cases}
      \begin{pmatrix}
       a_e(k) & 0 \\ 0 & \frac{1}{a_e(k)}
      \end{pmatrix}
 & \Im k >0,\\
\begin{pmatrix}
       \frac{1}{\overline{a_e(\bar k)}} & 0 \\ 0 & \overline{a_e(\bar k)}
      \end{pmatrix}, & \Im k < 0
\end{cases}
\end{equation}
and
\begin{equation}\label{P}
 \bar P(t)=\frac{1}{\Delta(t)} \begin{pmatrix}
 \hat M_{11}(0,t,-\ii{h}) & \hat M_{12}(0,t,\ii{h}) \\
\hat M_{21}(0,t,-\ii{h}) & \hat M_{22}(0,t,\ii{h})
                 \end{pmatrix}
\end{equation}
with 
\begin{eqnarray*}
\Delta(t) &=& \hat M_{11}(0,t,-\ii{h})\hat M_{22}(0,t,\ii{h})-\hat M_{12}(0,t,\ii{h})\hat M_{21}(0,t,-\ii{h} )\\
&=& |\hat M_{11}(0,t,-\ii{h})|^2 +|\hat M_{21}(0,t,-\ii{h} )|^2 >0.
\end{eqnarray*}
\end{lem}
\noindent Proof.
Define $M(t,k)$ by 
\begin{equation}\label{M-b}
 M(t,k)={N}(t,k)\overline{\hat M(0,t,-\bar k)}C(k),
\end{equation}
where $N(t,k)$ is some rational (in $k$) function with poles outside $\mathbb R$. 
Then the symmetries (\ref{J-sym}) and (\ref{sym-g}) imply that 
$M(t,k)$ satisfies the jump condition $M_+(t,k)=M_-(t,k)\hat J(0,t,k)$.
Now notice that the choice
\begin{equation}\label{E}
{N}(t,k)=\sigma_1 P(t)\begin{pmatrix}
                              \frac{1}{k-\ii{h}} & 0 \\ 0 &  \frac{1}{k+\ii{h}}
                             \end{pmatrix}
P^{-1}(t),
\end{equation}
with $P$ as in (\ref{P}) (for the details, see \cite{IS12}),
implies the r.h.s. of (\ref{M-b}) 
is nonsingular at $k=\pm\ii{h}$ and approaches
 the identity matrix as $k\to\infty$.
Then, checking the residue conditions (if any), we conclude that  
$M(t,k)=\hat M(0,t,k)$,
which eventually can be written as (\ref{M-sym}).

\bigskip

\begin{rem} The above arguments can actually  prove the general symmetry formula
that is valid for all $x$ and $t$:
$$
 \overline{\hat M(-x,t,-\bar k)} = \sigma_1 \bar P(t) \begin{pmatrix}
                              \frac{1}{k-\ii{h}} & 0 \\ 0 &  \frac{1}{k+\ii{h}}
                             \end{pmatrix}
\bar P^{-1}(t)\hat M(x,t,k)\begin{pmatrix}
 k-\ii{h} & 0 \\ 0 & k+\ii{h}
\end{pmatrix} D(k) \sigma_1.
$$
\end{rem}

Now we show  how the symmetry relation (\ref{M-sym}) alone  can be used to establish the Robin
boundary condition for $u(x,t)$. 
The arguments are exactly as in the case of decaying initial data \cite{IS12};
for a self-contained presentation, we repeat them here.

We first evaluate, using   
(\ref{M-sym}), the entries $\hat M_{11}(0,t,-\ii{h})$ and 
$\hat M_{21}(0,t,-\ii{h})$. We have:
\begin{equation}\label{11}
\begin{aligned} 
\overline{\hat M_{11}(0,t,-\ii{h})} & = \lim_{k\to -\ii{h}}\left(
\bar P(t) \begin{pmatrix}
                              \frac{1}{k-\ii{h}} & 0 \\ 0 &  \frac{1}{k+\ii{h}}
                             \end{pmatrix}
\bar P^{-1}(t)\hat M(0,t,k)\begin{pmatrix}
 k-\ii{h} & 0 \\ 0 & k+\ii{h}
\end{pmatrix} D(k)\right)_{22} \\
& = \bar P_{22}(t)\left(\bar P^{-1}(t)\hat M(0,t,-\ii{h})\right)_{22} d_2(-\ii{h}) = 
\frac{1}{\Delta(t)}\hat M_{22}(0,t,\ii{h})d_2(-\ii{h}) \\
& = \overline{\hat M_{11}(0,t,-\ii{h})}\frac{d_2(-\ii{h})}{\Delta(t)},
\end{aligned}
\end{equation}
where we have used the basic symmetry (\ref{sym}).
Similarly, we have 
\begin{equation}\label{21}
\begin{aligned} 
\overline{\hat M_{21}(0,t,-\ii{h})} & = \lim_{k\to -\ii{h}}\left(
\bar P(t) \begin{pmatrix}
                              \frac{1}{k-\ii{h}} & 0 \\ 0 &  \frac{1}{k+\ii{h}}
                             \end{pmatrix}
\bar P^{-1}(t)\hat M(0,t,k)\begin{pmatrix}
 k-\ii{h} & 0 \\ 0 & k+\ii{h}
\end{pmatrix} D(k)\right)_{12} \\
& = \bar P_{12}(t)\left(\bar P^{-1}(t)\hat M(0,t,-\ii{h})\right)_{22} d_2(-\ii{h}) = 
\frac{1}{\Delta(t)}\hat M_{12}(0,t,\ii{h})d_2(-\ii{h}) \\
& = -\overline{\hat M_{21}(0,t,-\ii{h})}\frac{d_2(-\ii{h})}{\Delta(t)}.
\end{aligned}
\end{equation}

Comparing (\ref{11}) and (\ref{21}) gives
\begin{equation}\label{m-0}
\overline{\hat M_{11}(0,t,-\ii{h})}(1-\theta)=0 \ \ \text{and}\ \ 
\overline{\hat M_{21}(0,t,-\ii{h})}(1+\theta)=0,
\end{equation}
where
\begin{equation}\label{theta}
\theta = \frac{d_2(-\ii{h})}{\Delta},
\end{equation}
which implies that either $\hat M_{11}(0,t,-\ii{h})=0$ or 
$\hat M_{21}(0,t,-\ii{h})=0$ for all $t\ge 0$.

Now recall that $\Psi(t,k):=(\hat M_{11}(0,t, k)\ee^{-2\ii k^2 t}, 
\hat M_{21}(0,t, k)\ee^{-2\ii k^2 t})^T$
satisfies the differential equation (\ref{Lax2}) with $u=u(0,t)$ and $u_x=u_x(0,t)$,
i.e.
\begin{equation}\label{dif}
\begin{aligned} 
\frac{d\Psi_1}{dt} + 2\ii k^2 \Psi_1 & = \ii|u(0,t)|^2\Psi_1+ (2ku(0,t)+\ii u_x(0,t))\Psi_2, \\
\frac{d\Psi_2}{dt} - 2\ii k^2 \Psi_2 & = -\ii|u(0,t)|^2\Psi_2+ (-2k\bar u(0,t)+\ii \bar u_x(0,t))\Psi_1.
\end{aligned}
\end{equation}
From (\ref{dif}) it follows that 
 if $\Psi_1(t,-\ii{h})=0$ for all $t\ge 0$ then $-2\ii{h} u(0,t)+\ii u_x(0,t)\equiv 0$,
or $u_x(0,t)-2{h} u(0,t)\equiv 0$;
 if $\Psi_2(t,-\ii{h})=0$ then $2\ii{h} \bar u(0,t)+\ii \bar u_x(0,t)\equiv 0$,
or $u_x(0,t)+2{h} u(0,t)\equiv 0$. Observe that, according to (\ref{be}), ${h}$ 
can be either $q/2$ or $-q/2$.
But since the initial data satisfy the boundary condition $u_x(0,0)+qu(0,0)=0$,
by continuity it follows that this condition holds for all $t$.

A closer look at (\ref{m-0}) and (\ref{theta}) reveals that one can specify precisely whether
(a) $\hat M_{11}(0,t,-\ii{h})=0$ or 
(b) $\hat M_{21}(0,t,-\ii{h})=0$ occurs, depending on the sign of $q$ and the properties 
of $a(\ii|q|/2)$ and $b(\ii|q|/2)$. Indeed, since 
$\Delta=|\hat M_{11}(-\ii{h})|^2+|\hat M_{21}(-\ii{h})|^2>0$,
the choice between (a) and (b) is determined by the sign of $d_2(-\ii{h})$.
According to (\ref{a-e}) and (\ref{be}), one can distinguish four cases.

(i) If $q>0$ and $b(\ii q/2)=0$, then ${h}=\frac{q}{2}>0$ and thus (see (\ref{D}))
$d_2(-\ii{h})=\overline{a_e(\frac{\ii q}{2})}$. In turn, from (\ref{a-e}) it follows that 
in this case, $\overline{a_e(\frac{\ii q}{2})}=\left|a(\frac{\ii q}{2})\right|^2>0$ and thus
$1+\frac{d_2(-\ii{h})}{\Delta}>0$, which implies (see(\ref{m-0})) that $\hat M_{21}(0,t,-\ii{h})=0$.

(ii) If $q<0$ and $a(-\ii q/2)\ne 0$, then ${h}=\frac{q}{2}<0$ and thus 
$d_2(-\ii{h})=\left(a_e(-\frac{\ii q}{2})\right)^{-1} = \left|a(-\frac{\ii q}{2})\right|^{-2}>0$.
Hence, in this case one also has $1+\frac{d_2(-\ii{h})}{\Delta}>0$
and thus $\hat M_{21}(0,t,-\ii{h})=0$.

(iii) If $q<0$ and $a(-\ii q/2)=0$, then ${h}=-\frac{q}{2}>0$ and thus
$d_2(-\ii{h})= \overline{a_e(-\frac{\ii q}{2})} =-\left|b(-\frac{\ii q}{2})\right|^{2}<0 $.
Hence, in this case $1-\frac{d_2(-\ii{h})}{\Delta}>0$, which implies that $\hat M_{11}(0,t,-\ii{h})=0$.

(iv) If $q>0$ and $b(\ii q/2)\ne 0$, then ${h}=-\frac{q}{2}<0$ and thus
$d_2(-\ii{h})=\left(a_e(\frac{\ii q}{2})\right)^{-1} = -\left|b(\frac{\ii q}{2})\right|^{-2}<0$.
Hence, in this case one also has $1-\frac{d_2(-\ii{h})}{\Delta}>0$, which 
implies that $\hat M_{11}(0,t,-\ii{h})=0$.

Summarizing, we see that $\hat M_{21}(0,t,-\ii{h})=0$ corresponds to ${h}=\frac{q}{2}$
while $\hat M_{11}(0,t,-\ii{h})=0$ corresponds to ${h}=-\frac{q}{2}$, which is indeed consistent 
with the fact that (\ref{dif}) implies $u_x(0,t)+qu(0,t)=0$ for all $t$.

\section{Continuation of the initial data}
\setcounter{equation}{0}

\begin{thm}\label{u0-}
Let $\hat M(x,t,k)$ be the solution of the RH problem 
with the jump matrices defined in terms of $q$ and $a(k)$ and $b(k)$ from (\ref{a-b})
via (\ref{r-e})--(\ref{j-0-h}).
Let $u(x,t)$ be determined by $\hat M(x,t,k)$ via (\ref{u-h}).
Then:
\begin{itemize}
	\item $u(x,t)$ solves the NLS equation for $-\infty<x<\infty$, $t>0$;
	\item $u_x(0,t)+q u(0,t)=0$ for $t>0$;
	\item $u(x,0)=\alpha\ee^{-2\ii\beta x}$ for $x>0$;
	\item $u(x,0) = \alpha \ee^{\ii\phi}\ee^{2\ii\beta x}$ for $x<0$,
\end{itemize}
where
\begin{equation}\label{u0}
\phi = \frac{-i}{\pi}\int_{\gamma_1}
\frac{\log(2\xi- \ii|q|)-\log(2\xi+ \ii|q|)}{(\sqrt{(\xi+\beta)^2+\alpha^2})_+}\dd\xi.
\end{equation}

\end{thm}

\noindent Proof. In view of Theorem \ref{main-thm}, the only item to be proved is the formula
$u(x,0) = \alpha \ee^{\ii\phi}\ee^{2\ii\beta x}$
giving the continuation of the initial data to the left half-line.
We will prove it via a series of transformations of the original RH problem
in the spirit of the nonlinear steepest descent method \cite{DZ93}.

First, let us express 
 the jump matrix $\hat J_0(k)$ in (\ref{M-h-jump})
in terms of $a_e(k)$ and $b_e(k)$ determined by (\ref{a-e}) and (\ref{b-e}).
\begin{lem}\label{Ga-a-b-lem}
The off-diagonal entries in (\ref{J0-ga1}), $k\in \gamma_1$, can be written as follows:
\begin{equation}\label{Ga-a-b}
\tilde \Gamma_+(k) - \tilde \Gamma_-(k) = \begin{cases}
\frac{2k+\ii q}{2k-\ii q}\frac{\ii}{a_{e+}(k)a_{e-}(k)}, & \text{case 1} \\
\frac{2k-\ii q}{2k+\ii q}\frac{\ii}{a_{e+}(k)a_{e-}(k)}, & \text{case 2}
\end{cases}.
\end{equation}
\end{lem}
The proof follows the direct calculation using the definition of $\tilde \Gamma$ and 
the identity (\ref{ga1-1}) from Lemma \ref{a-b-lim}.

Thus $\hat J_0(k)$ can be written as follows:
\begin{equation}\label{J0-new}
\hat J_0(k) = \begin{cases}
\begin{pmatrix}
\frac{a_{e-}(k)}{a_{e+}(k)} & \ii \\ 0 & \frac{a_{e+}(k)}{a_{e-}(k)}
\end{pmatrix}, & k\in \gamma\cap {\mathbb C}_+ \\[5mm]
\begin{pmatrix}
1 & 0 \\
\frac{2k\pm\ii q}{2k\mp\ii q}\frac{\ii}{a_{e+}(k)a_{e-}(k)} & 1 
\end{pmatrix}, & k\in \gamma_1\cap {\mathbb C}_+ \\[5mm]
\begin{pmatrix}
1 & \bar r_e(k) \\
0 & 1 
\end{pmatrix}
\begin{pmatrix}
1 & 0 \\
r_e(k) & 1 
\end{pmatrix}, & k\in \mathbb R
\end{cases}
\end{equation}
where $r_e=\frac{\bar b_e}{a_e}$, 
and 
\begin{equation}\label{j-sym}
\hat J_0(k) = \begin{pmatrix}
0 & 1 \\
-1 & 0 
\end{pmatrix}\overline{\hat J_0(\bar k)}\begin{pmatrix}
0 & -1 \\
1 & 0 
\end{pmatrix}
\end{equation}
for the parts of the contour in ${\mathbb C}_-$, i.e., for $k\in (\gamma\cup\gamma_1)\cap {\mathbb C}_-$.

Now define $M^{(1)}$ by
\begin{equation}\label{M1}
M^{(1)}(x,t,k) = \begin{cases}
\hat M(x,t,k)\begin{pmatrix}
a_e(k) & 0 \\
0 & a_e^{-1}(k) 
\end{pmatrix}, & k\in {\mathbb C}_+ \\[5mm]
\hat M(x,t,k)\begin{pmatrix}
\overline{a_e^{-1}(\bar k)} & 0 \\
0 & \overline{a_e(\bar k) }
\end{pmatrix}, & k\in {\mathbb C}_- 
\end{cases}
\end{equation}
Then the jump condition for $M^{(1)}(x,0,k)$ is as follows:
$$
M^{(1)}_+(x,0,k) =M^{(1)}_-(x,0,k)J^{(1)}(x,0,k),
$$
where 
\begin{equation}\label{J1}
J^{(1)}(x,0,k) = \begin{cases}
\begin{pmatrix}
1 & \frac{\ii\ee^{-2\ii k x }}{a_{e-}(k)a_{e+}(k)} \\ 0 & 1
\end{pmatrix}, & k\in \gamma\cap {\mathbb C}_+ \\[5mm]
\begin{pmatrix}
\frac{a_{e+}(k)}{a_{e-}(k)} & 0 \\[3mm]
\frac{2k\pm\ii q}{2k\mp\ii q}\ii\ee^{2\ii k x } & \frac{a_{e-}(k)}{a_{e+}(k)} 
\end{pmatrix}, & k\in \gamma_1\cap {\mathbb C}_+ \\[7mm]
\begin{pmatrix}
1 & 0 \\
\bar{\tilde r}_e(k)\ee^{2\ii k x } & 1 
\end{pmatrix}
\begin{pmatrix}
1 & \tilde r_e(k)\ee^{-2\ii k x } \\
0 & 1 
\end{pmatrix}, & k\in \mathbb R
\end{cases}
\end{equation}
where $\tilde r_e=\frac{b_e}{a_e}$, 
and
\begin{equation}\label{j-sym-2}
J^{(1)}(x,0,k) = \begin{pmatrix}
0 & 1 \\
-1 & 0 
\end{pmatrix}\overline{J^{(1)}(x,0,\bar k)}\begin{pmatrix}
0 & -1 \\
1 & 0 
\end{pmatrix}, \qquad k\in (\gamma\cup\gamma_1)\cap {\mathbb C}_-.
\end{equation}

The next transformation aims at removing the growing (as $x\to -\infty$) exponential in the jump
matrix on $\gamma_1$ (the ``$g$-function'' step, by now standard in the method of nonlinear steepest descent, see, e.g., \cite{BV07, BKS11}).
Define $M^{(2)}$ by 
$$
M^{(2)}(x,0,k)=\ee^{-\ii\beta x \sigma_3} M^{(1)}(x,0,k) \ee^{(-\ii k x+\ii \hat X(k) x)\sigma_3},
$$
where (cf. (\ref{x-om-nu})) $\hat X(k) = \sqrt{(k+\beta)^2+\alpha^2}$ with the branch cut $\gamma_1$
and $\hat X(k)=k+\beta =o(1)$ as $k\to\infty$ (so that the large-$k$ behavior is preserved: 
$M^{(2)}\to E$ as $k\to\infty$).
Then $\hat X_+ + \hat X_- = 0$ for $k\in\gamma_1$, and the jump matrix $J^{(2)}$
in the jump relation $M^{(2)}_+(x,0,k) =M^{(2)}_-(x,0,k)J^{(2)}(x,0,k)$ takes the form 
\begin{equation}\label{J2}
J^{(2)}(x,0,k) = \begin{cases}
\begin{pmatrix}
1 & \frac{\ii\ee^{-2\ii \hat X(k) x }}{a_{e-}(k)a_{e+}(k)} \\ 0 & 1
\end{pmatrix}, & k\in \gamma\cap {\mathbb C}_+ \\[5mm]
\begin{pmatrix}
\frac{a_{e+}(k)}{a_{e-}(k)}\ee^{\ii (\hat X_+(k)-\hat X_-(k)) x } & 0 \\[3mm]
\frac{2k\pm\ii q}{2k\mp\ii q}\ii & \frac{a_{e-}(k)}{a_{e+}(k)} \ee^{-\ii (\hat X_+(k)-\hat X_-(k)) x }
\end{pmatrix}, & k\in \gamma_1\cap {\mathbb C}_+ \\[7mm]
\begin{pmatrix}
1 & 0 \\
\bar{\tilde r}_e(k)\ee^{2\ii \hat X(k) x } & 1 
\end{pmatrix}
\begin{pmatrix}
1 & \tilde r_e(k)\ee^{-2\ii \hat X(k) x } \\
0 & 1 
\end{pmatrix}, & k\in \mathbb R
\end{cases}
\end{equation}
and 
$$
J^{(2)}(x,0,k) = \begin{pmatrix}
0 & 1 \\
-1 & 0 
\end{pmatrix}\overline{J^{(2)}(x,0,\bar k)}\begin{pmatrix}
0 & -1 \\
1 & 0 
\end{pmatrix}, \qquad k\in (\gamma\cup\gamma_1)\cap {\mathbb C}_-.
$$

Finally, introduce $M^{(3)}$ by 
\begin{equation}\label{M3}
M^{(3)}(x,0,k) = \begin{cases}
\begin{pmatrix}
f(\infty) & 0 \\
0 & f^{-1}(\infty)
\end{pmatrix}
M^{(2)}(x,0,k)\begin{pmatrix}
f^{-1}(k) & -f(k) h(k) \ee^{-2\ii \hat X(k) x} \\
0 & f(k)
\end{pmatrix}, & k\in {\mathbb C}_+ \\[5mm]
\begin{pmatrix}
\bar f^{-1}(\infty) & 0 \\
0 & \bar f(\infty)
\end{pmatrix}
M^{(2)}(x,0,k)\begin{pmatrix}
\bar f(\bar k) & 0 \\
 \bar f(\bar k) \bar h(\bar k) \ee^{2\ii \bar{\hat X}(\bar k) x} & {\bar f}^{-1}(\bar k)
\end{pmatrix}, & k\in {\mathbb C}_-
\end{cases}
\end{equation}
where the scalar functions $h(k)$ and $f(k)$ are to be determined.
Assuming that $h(k)$ and $f(k)$ are analytic for $k\in {\mathbb C}\setminus \{\gamma\cup\gamma_1\}$ and that $\bar f (\bar k) =f^{-1}(k)$, the jump matrix $J^{(3)}$
in $M^{(3)}_+(x,0,k) =M^{(3)}_-(x,0,k)J^{(3)}(x,0,k)$ take the form:

\begin{itemize}
	\item for $k\in \gamma_1\cap {\mathbb C}_+$: 
\begin{equation}\label{j3-1}
	J^{(3)}(x,0,k) = \begin{pmatrix}
	f_+^{-1}f_-\left(\frac{a_{e+}}{a_{e-}}+h_- \frac{2k\pm\ii q}{2k\mp\ii q}\ii\right)
	\ee^{-2\ii \hat X_- x} &
	f_-f_+\left(-h_+ \left(\frac{a_{e+}}{a_{e-}}+h_- \frac{2k\pm\ii q}{2k\mp\ii q}\ii\right)
	+h_- \frac{a_{e-}}{a_{e+}}\right) \\
	f_+^{-1}f_-^{-1}\frac{2k\pm\ii q}{2k\mp\ii q}\ii &
	f_-^{-1}f_+\left(\frac{a_{e-}}{a_{e+}}-h_+ \frac{2k\pm\ii q}{2k\mp\ii q}\ii\right)
	\ee^{-2\ii \hat X_+ x}
\end{pmatrix}
\end{equation}
\item for $k\in \gamma\cap {\mathbb C}_+$: 
\begin{equation}\label{j3-2}
J^{(3)}(x,0,k) = \begin{pmatrix}
1 & \left(\frac{\ii}{a_{e+}(k)a_{e-}(k)} +h_+(k)-h_-(k)\right)f_-(k)f_+(k)\ee^{-2\ii\hat X(k) x} \\
0 & 1 
\end{pmatrix}
\end{equation}
\item
for $k\in \gamma\cap {\mathbb C}_-$ and $k\in \gamma_1\cap {\mathbb C}_-$:
$\hat J^{(3)}(k) = \begin{pmatrix}
0 & 1 \\
-1 & 0 
\end{pmatrix}\overline{\hat J^{(3)}(\bar k)}\begin{pmatrix}
0 & -1 \\
1 & 0 
\end{pmatrix}$ 
\item
for $k\in \mathbb R$:
\begin{equation}\label{j3-3}
J^{(3)}(x,0,k) = \begin{pmatrix}
1 & 0 \\
(\bar{\tilde r}_e(k)-\bar h(k))f^{-2}(k)\ee^{2\ii \hat X(k) x } & 1 
\end{pmatrix}
\begin{pmatrix}
1 & (\tilde r_e(k)-h(k))f^2(k)\ee^{-2\ii \hat X(k) x } \\
0 & 1 
\end{pmatrix}
\end{equation}
\end{itemize}

\begin{lem}\label{J3-simple}
Setting $h(k):=\tilde r_e(k)$, the jump matrix $J^{(3)}(x,0,k)$ simplifies to 
$$
J^{(3)}(x,0,k) = \begin{pmatrix}
	0 &
	f_-(k)f_+(k)\frac{2k\mp\ii q}{2k\pm\ii q}\ii\\
	f_+^{-1}(k)f_-^{-1}(k)\frac{2k\pm\ii q}{2k\mp\ii q}\ii &
	0
\end{pmatrix}, \qquad k\in \gamma_1
$$
(so that $J^{(3)}\equiv E$ for $k\in \gamma$ and for $k\in \mathbb R$).
Moreover, $M^{(3)}(x,0,k)$ is a piecewise analytic function (having no poles). 

\end{lem}

Indeed, in the case of $a(k)$ and $b(k)$ given by (\ref{a-b}), 
$b_e(k)$ is analytic (similarly to $a_e(k)$) in $k\in {\mathbb C}\setminus \{\gamma\cup\gamma_1\}$; moreover, direct calculations give
$$
\frac{b_{e+}(k)}{a_{e+}(k)} = -\frac{a_{e-}(k)}{a_{e+}(k)}\frac{2k\mp\ii q}{2k\pm\ii q}\ii, \qquad
\frac{b_{e-}(k)}{a_{e-}(k)} = \frac{a_{e+}(k)}{a_{e-}(k)}\frac{2k\mp\ii q}{2k\pm\ii q}\ii
$$
and thus $J^{(3)}$ takes the required form for $k\in\gamma_1\cap {\mathbb C}_+$.
On the other hand, for $k\in\gamma$ one has 
$$
\frac{b_{e-}(k)}{a_{e-}(k)} - \frac{b_{e+}(k)}{a_{e+}(k)} = \frac{\ii}{a_{e+}(k)a_{e-}(k)}
$$
and thus $J^{(3)}$ reduces to $E$ on $k\in \gamma\cap {\mathbb C}_+$.

The fact that $M^{(3)}(x,0,k)$ has no poles at the zeros of $a_e(k)$ (if any)
can be checked directly.

\medskip

Now let us specify $f(k)$ as a solution of the following factorization problem:
\begin{itemize}
	\item 
	$f(k)$ is analytic in $\mathbb C\setminus \gamma_1$
	\item
		$f(k)$ is bounded as $k\to\infty$
		\item
		the limiting values $f_\pm$ of $f$ on $\gamma_1$ are related as follows:
		$
		f_+(k)f_-(k)=\frac{2k\pm\ii q}{2k\mp\ii q}
		$.
\end{itemize}
Noting that $\tilde f:=\frac{\log f}{\hat X}$ has to satisfy the jump condition
$\tilde f_+-\tilde f_-=\frac{\log \eta}{\hat X_+}$ and $\tilde f\to 0$ as $k\to\infty$,
the Cauchy-type integral formula for $f$ follows:
$$
f(k)=\exp\left\{\frac{\hat X(k)}{2\pi}\int_{\gamma_1}\frac{\log\eta(\xi)}{\hat X_+(\xi)(\xi-k)}d\xi\right\},
$$
where $\eta(k):=\frac{2k\pm\ii q}{2k\mp\ii q}$. 
Particularly, we have
\begin{equation}\label{f-inf}
f(\infty) = \exp\left\{-\frac{1}{2\pi}\int_{\gamma_1}\frac{\log\eta(\xi)}{\hat X_+(\xi)}d\xi\right\}.
\end{equation}

With $f$ determined in this way,  $J^{(3)}(x,0,k)$ reduces to a matrix with
constant entries and thus the RH problem for $M^{(3)}$ reduces to 
\begin{itemize}
	\item 
\begin{equation}\label{j3-mod}
M^{(3)}_+(x,0,k) = M^{(3)}_-(x,0,k)\begin{pmatrix}
	0 & \ii \\ \ii & 0
\end{pmatrix}, \qquad k\in \gamma_1
\end{equation}
\item
$M^{(3)}(x,0,k)\to E$ as $k\to\infty$
\end{itemize}

The unique solution of this RH problem, having singularities at the end points of 
$\gamma_1$ of order not greater than $1/2$, is given explicitly:
$$
M^{(3)}_+(x,0,k) = \frac{1}{2}\begin{pmatrix}
	\varkappa(k) + \varkappa(k)^{-1} & \varkappa(k) - \varkappa(k)^{-1}  \\ 
	\varkappa(k) - \varkappa(k)^{-1}  & \varkappa(k) + \varkappa(k)^{-1} 
\end{pmatrix}
$$
with $\varkappa(k)=\left(\frac{k+\beta -\ii\alpha}{k+\beta +\ii\alpha}\right)^{\frac{1}{4}}$.

Going back to $\hat M$ we have:
\begin{eqnarray*}
\hat M(x,0,k) &=& \ee^{\ii\beta x \sigma_3} \begin{pmatrix}
	f^{-1}(\infty) & 0 \\ 0 & f(\infty)
\end{pmatrix}
M^{(3)}_+(x,0,k) 
\begin{pmatrix}
	f(k) & f(k)h(k)\ee^{-2\ii\hat X(k)x} \\ 0 & f^{-1}(k)
\end{pmatrix} \nonumber\\
&& \times \ee^{(\ii k x - \ii \hat X(k) x)\sigma_3}
\begin{pmatrix}
	a_e^{-1}(k) & 0 \\ 0 & a_e(k)
\end{pmatrix}
\end{eqnarray*}
Taking into account that $\varkappa(k)=1+\frac{\alpha}{2\ii k} + O(\frac{1}{k^2})$ as $k\to\infty$ we conclude that for all $x<0$,
$$
u(x,0)=\lim_{k\to\infty} 2\ii k \hat M_{12}(x,0,k) = \alpha \ee^{2\ii\beta x} f^{-2}(\infty).
$$
Finally we notice that in the case of exact exponential initial data,
$u_0(x)=\alpha\ee^{-2\ii\beta x}$, case 1 is realized with $q<0$ whereas case 2 is realized with $q>0$ and thus the additional phase factor, coming from $f^{-2}(\infty)$,
see (\ref{f-inf}),
takes the form as in (\ref{u0}).

\hfill $\Box$

\begin{rem}
Similarly to Theorem \ref{main-thm}, a more general result holds: the RH problem constructed 
from the spectral functions determined by $u(x,0)=u_0(x)$, $x>0$ with $u_0(x)\to \alpha \ee^{-2\ii\beta x}$
as $x\to +\infty$ gives rise to $u(x,t)$ such that $u(x,0)\to \alpha \ee^{\ii\phi}\ee^{2\ii\beta x}$
as $x\to -\infty$. Similarly to the case of exact exponential initial data,
the proof is based on a sequence of transformations of the RH problem leading the 
the same model problem with the jump (\ref{j3-mod}).

A major difference, comparing to the case of exact exponential initial data, is that in passing from $M^{(2)}$ to 
$M^{(3)}$, one cannot expect to determine $h(k)$ as an analytic function in the whole half-plane $\mathbb C_+$ that allows getting rid of the diagonal terms in the jump matrix on 
$\gamma_1$, see (\ref{j3-1}), and simultaneously, reducing (\ref{j3-2}) and (\ref{j3-3})
 exactly to the identity matrix. Instead, lens-shaped domains are to be introduced 
containing $\gamma_1\cap
{\mathbb C}_+$,  $\gamma_1\cap
{\mathbb C}_-$ and  the rays $(-\infty, -\beta)$ 
and $(-\beta, \infty)$, together with appropriated functions, analytic in the lenses,
that approximate respectively $-\frac{a_{e-}(k)}{a_{e+}(k)}\frac{2k\mp\ii q}{2k\pm\ii q}\ii$, $\frac{a_{e+}(k)}{a_{e-}(k)}\frac{2k\mp\ii q}{2k\pm\ii q}\ii$ on $\gamma_1\cap
{\mathbb C}_+$ and $\frac{b_{e}(k)}{a_{e}(k)}$ on $\mathbb R$. Then $M^{(3)}$ is determined by $M^{(2)}$ multiplied, in each lens, by a particular triangular factor
similar to (\ref{j3-1})--(\ref{j3-3}), which leads to the jump matrix $J^{(3)}$
involving, apart from (\ref{j3-mod}) for $k\in\gamma_1$, the jump matrices on the boundaries 
of the lenses that approach the identity matrix as $x\to -\infty$. As a result, 
$u(x,0)$ obtained via the solution of the RH problem does not coincide, 
for all $x<0$, with $\alpha\ee^{2\ii\beta x}$
 as in Theorem \ref{u0-} but approaches this exponential as $x\to-\infty$.
The “opening of lenses” idea is also standard in the method of steepest descent; 
see \cite{BKS11, BV07} in the context of
focusing NLS with steplike initial data.

\end{rem}

\section{Large-time asymptotics: rarefaction case ($\beta<0$)}
\setcounter{equation}{0}

The RH problems in the cases $\beta>0$ and $\beta<0$ differ only 
 in the positions of $\gamma$ and $\gamma_1$.
 This difference, being irrelevant in proving that the solution of the RH problem indeed
 satisfies the Robin boundary condition and the prescribed initial condition, 
 becomes important in the  analysis of the long-time behavior of the solution of the IBV problem.
 
 In the case $\beta>0$, $\gamma$ is located in the right half-plane and $\gamma_1$ is located
 in the left half-plane.
 The asymptotic analysis of the 
 RH problem  in this situation  was given (for $q=0$) in \cite{BV07}, where it was shown that 
 a shock-type large-time behavior occurs: 
one can specify $\xi_j=\xi_j(\alpha,\beta)$, $j=1,2$ such that
(i) in the sector $\left|\frac{x}{t}\right|<\xi_1$ of the $(x,t)$ plane
 (containing, in particular, the vertical rays corresponding to fixed $x$ and $t\to
 +\infty$), the large-time behavior is oscillatory and can be 
 described in terms of the single phase theta functions;
(ii) in the sectors
$\left|\frac{x}{t}\right|>\xi_2$, the asymptotics has a plane wave form;
(iii) in two transition sectors 
$\xi_1<\left|\frac{x}{t}\right|<\xi_2$, the asymptotics is expressed in terms 
of the two phase theta functions.
One can show that the presence of $q\ne 0$ does not change the location of the boundaries
of the sectors and  the qualitative behavior of the solution in the 
corresponding sectors, but it does break the symmetry with respect to the $t$-axis
by adding a factor similar to $\ee^{\ii\phi}$ in Theorem \ref{u0-}. 
 
In the case $\beta<0$, 
 $\gamma$ and $\gamma_1$ interchange their locations, which implies, in particular,  that in the sector $|x/t|\le 4|\beta|$ (particularly, for all fixed $x$),
 the exponential factors
 in the jump matrices on $\gamma$ and $\gamma_1$, see (\ref{J0-ga}), (\ref{J0-ga1}),
(\ref{M-h-jump}),
 are all decaying, as $t\to \infty$.
 Consequently, 
in this sector,  the contribution 
 of the jumps across $\gamma$ and $\gamma_1$ is vanishing as $t\to \infty$
 and thus $u(x,t)$ decays to $0$ unless there are $k_j$, $j=1,2,\dots$
such that $|\Re k_j|<|\beta|$ and $a_e(k_j)=0$, which generate 
 residue conditions for the RH problem. 
 
Specifically, consider  the case of pure exponential initial conditions:
 $u(x,0)= \alpha \ee^{-2\ii\beta x}$ for $x>0$. 
Depending of the sign of $q$, we have two 
 situations:
 \begin{itemize}
 \item
 if $q<0$, then $a_e(k)$ has no zeros with $|\Re k|<|\beta|$.
 \item
 if $q>0$, then $a_e(k)$ has exactly one zero, which is  located on the imaginary axis:
$a_e(\ii\zeta)=0$ with some $\zeta>0$.
 \end{itemize}
 
 Consequently, if $q<0$, then $u(x,t)$ is rapidly decaying as $t\to\infty$ in the sector $\left|\frac{x}{t}\right|<4|\beta|$.
As for the case 
 $q>0$, in this sector $u(x,t)$ behaves like a breather.
In order to demonstrate this, we first notice that the residue conditions (\ref{res})
can be written in terms of $a_e(k)$ and $b_e(k)$ only. Indeed, in the case of pure
exponential initial data, $b_e(k)$ is analytic in 
${\mathbb C}\setminus (\gamma\cup\gamma_1)$ and, using the symmetries in (\ref{a-b}),
can be written as (cf. (\ref{b-e}))
\begin{equation}\label{b-e-1}
 b_e(k):=\begin{cases}
b(k)\overline{a(-\bar k)} - \frac{2k-\ii q}{2k+\ii q}\overline{b(-\bar k)}a(k), &
	\text{if }\  q<0,  \\
\frac{2k+\ii q}{2k-\ii q}b(k)\overline{a(-\bar k)} - \overline{b(-\bar k)}a(k), &
	\text{if }\  q>0.
        \end{cases}
\end{equation}
Then, using the definition (\ref{a-e}) of $a_e(k)$, it is readily seen that 
if $k^*$ is such that $a_e(k^*)=0$, then (cf. (\ref{res}))
$$
\frac{1}{b_e(k^*)} = \frac{\overline{b(-\bar k^*)}}{a(k^*)}\chi(k^*)
$$
with $\chi(k)$ defined in (\ref{chi}),
and thus the residue conditions can be written, in both cases, as 
\begin{eqnarray}\label{res1}
\R{Res}_{k_j}\hat M^{(1)}(x,t,k) &=& 
\frac{\ee^{2\ii\Theta(x,t,k_j)}}{\dot a_e(k_j)b_e(k_j)}\chi(k_j)\hat M^{(2)}(x,t,k_j),\nonumber\\
\R{Res}_{\bar k_j}\hat M^{(2)}(x,t,k) &=&
\frac{\ee^{-2\ii\bar\Theta(x,t,k_j)}}{\overline{\dot a_e(k_j)}\overline{b_e(k_j)}}\bar \chi(k_j)\hat M^{(1)}(x,t,\bar k_j).
\end{eqnarray}
Then,  
applying the procedure of reducing a (singular) RH problem with residue conditions to a regular
RH problem (see, e.g., \cite{FT}), one arrives at the following 
 theorem.
\begin{thm}\label{as-breezer}
Consider the initial boundary value problem (\ref{ibv}), where 
$u_0(x)=\alpha\ee^{-2\ii\beta x}$ with $\beta<0$ and $q>0$.
Then for $0\le \frac{x}{t}<4|\beta|$, the solution $u(x,t)$ has the asymptotics
$$
u(x,t)=\tilde u(x,t)\left(1+o(1)\right), \qquad t\to\infty,
$$
where 
$$
\tilde u(x,t) = \ii \ee^{-\ii\eta}\frac{\sqrt{2\omega}}{\cosh(\sqrt{2\omega} x +\phi_0)}
\ee^{2\ii\omega t}.
$$
 Here the parameters are defined as follows:
\begin{itemize}
	\item 
	$\omega = 2\zeta^2$, where $\zeta>0$ is 
	the (unique) solution of the equation
	$a_e(\ii\zeta)=0$, and 
	$a_e(k)$ is determined in (\ref{a-e}) with $a(k)$ and $b(k)$ determined by (\ref{a-b})
	and (\ref{x-om-nu});
	\item
	$\eta=\arg(\gamma)$, where $\gamma=\frac{1}{\dot a_e(\ii\zeta)b_e(\ii\zeta)}$;
	\item
	$\phi_0=\log \frac{2\zeta}{|\gamma|}$.
\end{itemize}
In particular, at $x=0$, $u(0,t)$ and
$u_x(0,t)$
exhibit single exponential oscillations:
$$
u(0,t)=\tilde u_1(t)\left(1+o(1)\right) \quad\text{and}\quad u_x(0,t)=\tilde u_2(t)\left(1+o(1)\right),
\qquad t\to\infty,
$$
where
$$
\tilde u_1(t)=\ee^{-\ii(\eta-\frac{\pi}{2})} \frac{\sqrt{2\omega}}{\cosh(\phi_0)}\ee^{2\ii\omega t},
$$
$$
\tilde u_2(t)=-\ee^{-\ii(\eta-\frac{\pi}{2})} \frac{2\omega\sinh(\phi_0)}{\cosh^2(\phi_0)}
\ee^{2\ii\omega t}.
$$
\end{thm}

\section*{Acknowledgments}

D.Sh. gratefully acknowledges the support of the Austrian Science Fund (FWF) under Grant No.\ Y330 and the
hospitality of the Archimedes Center for Modeling, Analysis and Computation (University of Crete), Laboratoire de Math\'{e}matiques Pures et Appliqu\'{e}es (Universit\'{e} du Littoral C\^{o}te d'Opale), and the University of Vienna,
where parts
of this research were done.

S.K. acknowledges the support of the European Union’s Seventh Framework Programme (FP7- REGPOT-2009-1) under grant 
agreement n 245749, as well as the ARISTEIA II grant n. 3964 from the Greek General Secretariat of Research and Technology.

$^1$
Department of Applied Mathematics, University of Crete \\
71409 Knossos-Heraklion, Crete, Greece \\
spyros@tem.uoc.gr

\medskip

$^2$B.~Verkin Institute for Low Temperature Physics and Engineering\\
47 Lenin Avenue, 61103 Kharkiv, Ukraine\\
shepelsky@yahoo.com

\medskip

$^3$Laboratoire de Math\'{e}matiques Pures et Appliqu\'{e}es, 
Universit\'{e} du Littoral C\^{o}te d'Opale,
50 rue Ferdinand Buisson - B.P. 699
62228 Calais, France\\
Lech.Zielinski@lmpa.univ-littoral.fr

\end{document}